%% file: SRP2016.tex
\documentclass[journal]{IEEEtran}
\usepackage{ifpdf}
\usepackage{cite}
\usepackage{color}
\usepackage{amsthm}
\usepackage{amsmath,amssymb}
\usepackage{algorithmic}
\usepackage{array}
\usepackage{mdwmath}
\usepackage{mdwtab}
\usepackage{graphicx}
\usepackage{lipsum}
\usepackage{epstopdf }
\usepackage{algorithm,algorithmic}
\usepackage{enumerate}
\usepackage{eqparbox}
\usepackage{url}
\usepackage{subfig}
\usepackage{amsbsy}

\input{mymath.tex}

\begin{document}
\title{\huge Online Ski Rental for ON/OFF Scheduling of \\Energy Harvesting Base Stations\vspace{-2mm}}
\author{\IEEEauthorblockN{Gilsoo~Lee$^{\dag}$,~Walid~Saad$^{\dag}$,~Mehdi~Bennis$^\ddag$,~Abolfazl~Mehbodniya$^{\S}$, and~Fumiyuki~Adachi$^{\S}$}\\
\IEEEauthorblockA{
\small $^{\dag}$ Wireless@VT, Department of Electrical and Computer Engineering, Virginia Tech, Blacksburg, VA, USA, \\ 
Emails: \protect\url{{gilsoolee, walids}@vt.edu}. \\
\small $^\ddag$ Centre for Wireless Communications, University of Oulu, Finland, Email: \url{bennis@ee.oulu.fi}.\\
\small $^\S$ Dept. of Communication Engineering, Graduate School of Engineering, Tohoku University, Sendai, Japan,\\ Emails:
\protect\url{mehbod@mobile.ecei.tohoku.ac.jp}, \protect\url{adachi@ecei.tohoku.ac.jp}.\vspace{-10mm}
}
\thanks{This research been supported by the U.S. National Science Foundation under Grant CNS-1460333 and by Towards Energy-Efficient Hyper-Dense Wireless Networks with Trillions of Devices, the Commissioned Research of National Institute of Information and Communications Technology (NICT), Japan, and the Academy of Finland CARMA project. 
A preliminary conference version \cite{lee2016online} of this work was presented at  IEEE ICC 2016.}
}
\maketitle

\begin{abstract} 
The co-existence of small cell base stations (SBSs) with conventional macrocell base station is a promising approach to boost the capacity and coverage of cellular networks. However, densifying the network with a viral deployment of SBSs can significantly increase energy consumption.  To reduce the reliance on unsustainable energy sources, one can adopt self-powered SBSs that rely solely on energy harvesting. Due to the uncertainty of energy arrival and the finite capacity of energy storage systems,  self-powered SBSs must smartly optimize their ON and OFF schedule. In this paper, the problem of ON/OFF scheduling of self-powered SBSs is studied, in the presence of energy harvesting uncertainty with the goal of minimizing the operational costs consisted of energy consumption and transmission delay of a network. For the original problem, we show an algorithm can solve the problem in the illustrative case. Then, to reduce the complexity of the original problem, an approximation is proposed. To solve  the approximated problem, a novel approach based on the \textit{ski rental framework}, a powerful online optimization tool, is proposed. Using this approach, each SBS can effectively decide on its ON/OFF schedule autonomously, without  any prior information on future energy arrivals. By using competitive analysis, a deterministic online algorithm (DOA) and a randomized online algorithm (ROA) are developed. The  ROA is then shown to achieve  the optimal competitive ratio in the approximation problem. Simulation results show that, compared to a baseline approach, the ROA can yield  performance gains reaching up to $15.6\%$ in terms of reduced total energy consumption of SBSs and up to $20.6\%$ in terms of per-SBS network delay  reduction. The results also  shed light on the fundamental aspects that impact the ON time of SBSs while demonstrating that the proposed  ROA can reduce up to $69.9\%$  the total cost compared to a baseline approach. \vspace{-2mm}
\end{abstract}


\section{Introduction}
Despite their promising potential for enhancing the capacity and coverage of cellular systems, small cell networks (SCNs) can also increase the overall power consumption of a cellular system since the access network and edge facilities take up to 83\% of mobiles' operator power consumption \cite{hwang2013holistic}. To this end,  enhancing the energy efficiency of dense SCNs has emerged as a major research challenge \cite{ashraf2011sleep}. In particular, there has been a recent significant interest, not only in minimizing energy consumption, but also in maximizing the use of green energy by deploying energy harvesting, self-powered base stations (BSs) that rely solely on renewable and clean energy for operation \cite{ansari2015gate}. Thus, deploying self-powered BSs is currently being demonstrated by various network operators. For instance, LG Uplus deploys solar-powered LTE BSs in mountain areas of South Korea \cite{lguplus}, and, also, a large solar-powered BS cluster is deployed in Tibet  by China Mobile \cite{bao2014solar}. Clearly, one can realize the vision of truly green cellular networks by deploying self-powered, energy harvesting small cell base stations (SBSs) that rely solely on renewable energy for their operation \cite{mao2015energy}. 

Recently, numerous  works have focused on the use of energy harvesting techniques in cellular networks \cite{han2014powering, han2013green, maghsudi2016distributed, maghsudi2016downlink, dhillon2014fundamentals, han2013optimizing, liu2015two, han2014provisioning, gong2014base, zhou2013sleep}. For instance, the work in \cite{han2014powering} overviews key design issues for adopting energy harvesting into cellular networks and propose energy harvesting-aware user association and BS sleep mode optimization problems. With regards to the user association problem in energy harvesting scenarios, the authors in \cite{han2013green} consider a model in which wireless BSs are powered by both grid power and green energy in energy harvesting heterogeneous cellular networks. For this model, the authors propose a user association scheme that minimizes the average traffic delay while maximizing the use of green energy. Furthermore, the authors in \cite{maghsudi2016distributed} propose a probabilistic framework to model energy harvesting and energy consumptions of BSs and investigate a distributed user association problem when BSs is powered by energy harvesting. Also, to study the problem of user association, in \cite{maghsudi2016downlink}, the authors considered a network in which the uncertainty of energy harvesting is modeled within a competitive market with the SBSs being the consumers who seek to maximize their utility function.

Reaping  the benefits of self-powered SBSs mandates effective and self-organizing ways to optimize the ON and OFF schedules of such SBSs, depending on uncertain and intermittent energy arrivals. Therefore, several recent works have focused on optimizing energy efficiency in energy harvesting systems by intelligently turning BSs ON and OFF \cite{dhillon2014fundamentals, han2013optimizing, liu2015two, han2014provisioning, gong2014base, zhou2013sleep}. For instance, the authors in \cite{dhillon2014fundamentals} provide a model to measure the performance of  heterogeneous networks with self-powered BSs. In \cite{han2013optimizing}, when BSs are powered by both a renewable source and the power grid, the authors propose an algorithm to maximize the utilization of green energy so that the grid power consumption can be minimized. Moreover, the work in \cite{liu2015two} develops a number of algorithms to minimize grid power consumption when considering hybrid-powered BSs. For solving a capital expenditure minimization problem, the authors in \cite{han2014provisioning} propose an ON/OFF scheduling method for self-powered BSs. The work in \cite{gong2014base} investigates the  problem of minimizing  grid power consumption and blocking probability by using statistical information for traffic and renewable energy. The authors in \cite{zhou2013sleep} study the optimal BS sleep policy based on dynamic programming with the statistical energy arrival information.

In this existing body of literature that addresses ON/OFF scheduling in energy harvesting networks \cite{han2013optimizing,liu2015two,han2014provisioning,gong2014base,zhou2013sleep},  it is generally assumed that statistical or complete information about the amount and arrival time of energy is  perfectly known. However, in practice, energy arrivals are largely intermittent and uncertain since they can stem from multiple sources. Moreover, turning SBSs ON and OFF based on every single energy arrival instance can lead to significant handovers and network stoppage times. Further, the existing works \cite{liu2015two, han2013green,han2014powering}, and \cite{gong2014base} on energy harvesting  networks often assume the presence of both smart grid and energy harvesting sources at every SBS.       
In contrast, here, we focus on cellular networks in which SBSs are completely self-powered and reliant on energy harvesting. In \cite{samarakoon2014dynamic} and \cite{samarakoon2016dynamic} the problem of ON/OFF scheduling of base stations is studied for a heterogeneous network using reinforcement learning. However, these works are focused on classical grid-powered networks and do not take into account the presence of energy harvesting in the system. 
Also, unlike the work in \cite{dhillon2014fundamentals} which focuses on the global performance analysis of self-powered SBSs, our goal is to develop self-organizing and online algorithms for optimizing the ON/OFF schedule of self-powered SBSs. 

The main contributions of this paper is to develop  a novel framework for optimizing the ON and OFF schedule of self-powered SBSs in a cellular network in which multiple  SBSs coexist with a macrocell base station (MBS). In particular, an optimization problem is formulated that seeks to minimize the operational cost that captures both the power and delay of the system by appropriately determining the SBSs ON and OFF scheduling, \emph{in the presence of complete uncertainty on the energy harvesting process}. We cast the problem as an online optimization and we analyze its properties. We show that, under an illustrative case, an algorithm achieves a competitive ratio, defined as the ratio of an online algorithm's to the optimal cost of an offline algorithm, of $2$. Then, to overcome the complexity of the original problem, an approximation is derived and shown to allow the decomposition of the original problem into a set of distributed online optimization problems that are run at each SBS. To solve the resulting per-SBS online optimization problem, a novel approach based on the \emph{ski rental problem}, a powerful online optimization tool \cite{lotker2008ski}, is proposed. In particular,  we present two schemes to solve the ski rental problem: a deterministic online algorithm (DOA) and a randomized online algorithm (ROA). On the one hand, the DOA is a benchmark scheme designed to turn each SBS OFF at a predetermined time so as to achieve a competitive ratio of $2$. On the other hand, the ROA enables the SBSs to make a decision according to a probability distribution, and it can achieve an optimal competitive ratio of $e/(e-1)$ which provides an upper bound for the approximated problem. The proposed algorithms  allow the SBSs to effectively decide on their ON/OFF schedule, without knowing any prior information on future energy arrivals. \emph{To the best of our knowledge, this is the first work that exploits the online ski rental problem for managing energy uncertainty in cellular systems with self-powered SBSs.} Simulation results show that the empirical competitive ratio of using the ROA to solve the original problem is $1.86$. This demonstrates that the  ROA achieves a reasonable performance gap compared to the ideal, offline optimal solution found by exhaustive search. Also, our results show that the ROA can decrease the total operational cost compared to the DOA and a baseline approach. Moreover,  the ROA can reduce total energy consumption of SBSs and per-SBS network delay compared to DOA or a baseline that turns SBSs ON during the same fixed period for all SBS.  This performance advantage is shown to reach up to $15.6\%$ and $11.4\%$ in reducing the energy consumption of a network relative to a baseline and the DOA, respectively.  The ROA  also decreases the delay per SBS up to $20.6\%$ and  $8.4\%$ relative to a baseline and the DOA, respectively. In particular, we observe that the ON time of each SBS is affected by various factors including the harvested energy and the power consumption of BSs. 

The rest of this paper is organized as follows. In Section \ref{sec:systemmodel}, the system model is presented.  
In Section \ref{sec:problemformulation}, we present the problem formulation. In Section \ref{sec:skirental}, we propose online algorithms based on the ski rental framework. In Section \ref{sec:numericalresults}, the performance of the proposed algorithm is demonstrated with using extensive simulations. Finally, conclusions are drawn  in Section \ref{sec:conclusion}.  \vspace{-4mm}

\section{System Model} \label{sec:systemmodel}

Consider the downlink of a two-tier heterogeneous small cell network in which an MBS is  located at the center of a service area. In this network, a set $\cJ$ of $J$ self-powered SBSs are deployed. Moreover, we define the set of all BSs as $\cB = \{0, 1, 2, \cdots, J\}$ where the MBS is indexed by $0$. We assume that the SBSs and the MBS will use different frequency bands and, therefore, the MBS and the SBSs will not interfere. In contrast, within the SBS tier, frequency bands may be reused and, as such, the SBS will interfere with one another. In this system, when activated, the SBSs can  offload traffic from the MBS, thus reducing the overall network congestion. A set $\cI$ of~$I$~UEs is randomly distributed in the coverage of the MBS where each UE can access either an SBS or MBS.   Each UE can be connected with only one of the BSs at a certain time $t$ within a period of $T$. 

\begin{figure}[t]
\centering 
\includegraphics[width=0.41\textwidth]{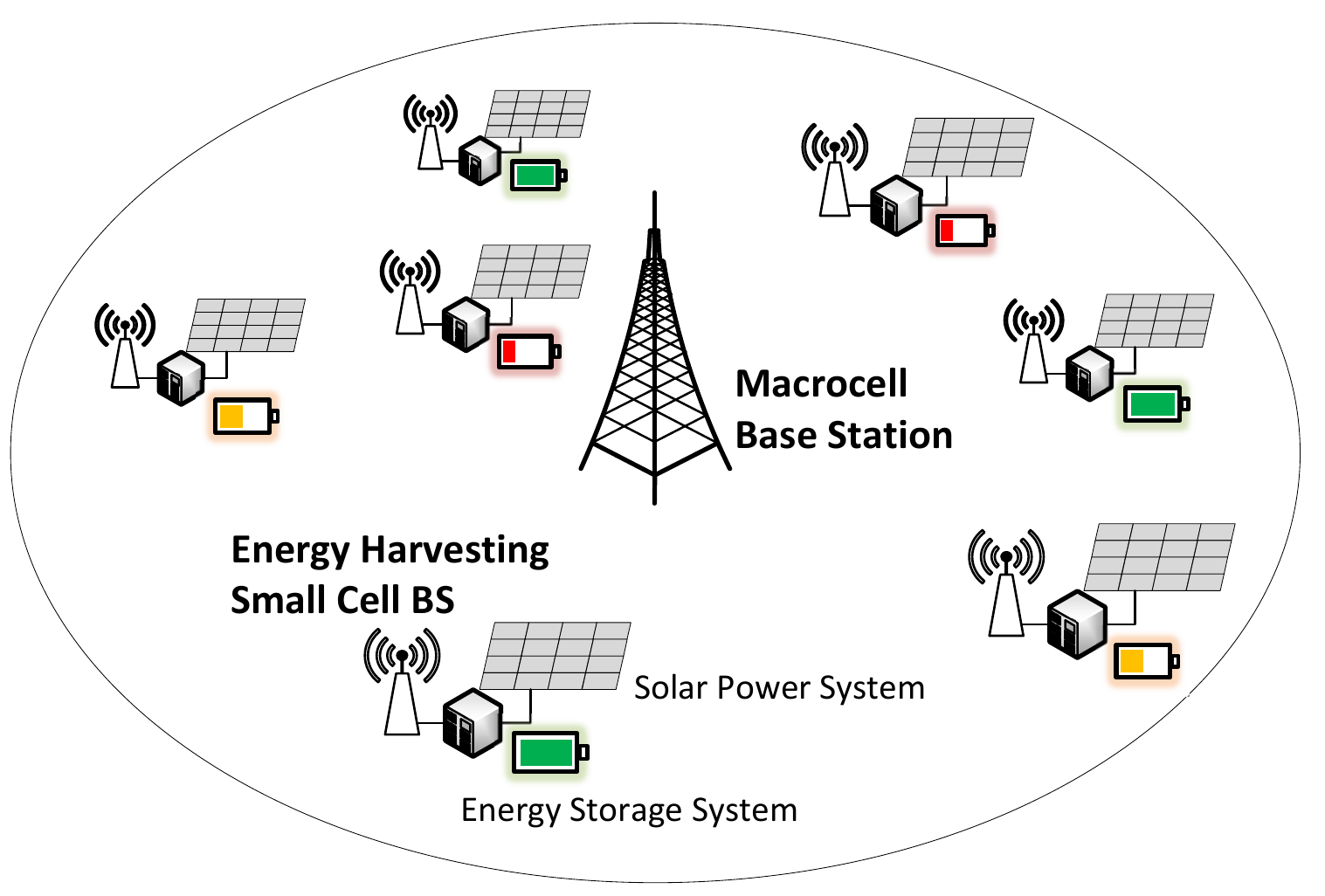}\vspace{-2mm}
\caption{\small System model of a heterogeneous deployment with self-powered SBSs.}
\label{fig:system}
\end{figure}

An illustration of our system model is shown in Fig.~\ref{fig:system}. In our considered system, while  the MBS is connected to the conventional power grid, SBSs are self-powered and rely exclusively on energy harvesting sources. In such case, the self-powered SBSs will operate as a means to boost capacity and to complement the existing grid powered MBS. For example,  SBSs can be equipped with solar panels   to procure energy for their operation, or, alternatively, they can use wireless power transfer from  MBS transmissions. Since the characteristics of the harvested energy can be highly dynamic, we do not make any specific assumption on the energy harvesting process. Thus, our model can accommodate any type of energy harvesting mechanism. To enhance the overall energy efficiency of the system, we assume that the SBSs can dynamically turn ON or OFF, depending on the network state, energy harvesting state, and other related parameters.  
To manage the  intermittent and uncertain nature of energy harvesting, energy storage systems (ESS) can be used. Energy harvesting is assumed to be done irrespective on whether an SBS is turned ON or OFF. Thus, an SBS will store energy in its ESS when it is turned OFF, and this stored energy  can be used when it is turned ON to service users. Also, when it is turned ON, an SBS can store the excess of harvested energy  if instantaneous harvested energy is enough to operate an SBS.  

At time $t$, the ON or OFF state of SBS $j$ is denoted by $\sigma_j(t)$ which is defined as follows: \vspace{-3mm}
\beq\label{eq:onoffstate}
			\sigma_j(t) = ~\left\{\begin{matrix}
			1,&{\text{if SBS }} j {\text{ is turned ON at time $t$}} ,  \label{eq:sigma_j}\\ 
			0,&{\text{otherwise}}.
			\end{matrix}\right.
\eeq
For the MBS, $\sigma_0(t)=1$ since the MBS is always turned ON. The set of switched-ON BSs at time $t$ is denoted by $\cB^{\textrm{on}}(t) = \{ j | \sigma_j(t)=1,  \forall j \in \cB\}$. Similarly, the set of switched-OFF BSs can be shown as $\cB^{\textrm{off}}(t)=\cB \setminus \cB^{\textrm{on}}(t)$. \vspace{-5mm}

\subsection{Network Performance}
We model the network performance between BS and UE. In the downlink, the signal to interference and noise ratio (SINR) between UE $i$ and SBS $j \in \cJ$ at time $t$ can be shown as
\beq\label{eq:sinr}
  \gamma_{ij}({\boldsymbol{\sigma}(t)}) = \frac{ P^\textrm{tx}_j \sigma_j(t) h_{ij} }{\sum_{j' \in \cB^{\textrm{on}} \setminus \{j\}} P^\textrm{tx}_{j'} \sigma_{j'}(t) h_{ij'} + \rho^2}, 
\eeq\vspace{-3mm}\\
where ${\boldsymbol{\sigma}(t)} = [\sigma_j(t)|\forall j \in \cJ]$, $h_{ij}$ is the channel gain between UE $i$ and SBS $j$, $P^\textrm{tx}_j$ is the transmit power of the connected SBS $j$, and  $\rho^2$ is the  noise power. If an UE is associated with an SBS, the UE can receive interference from the other SBSs.  On the other hand, when a UE is associated with the MBS, the UE does not experience any interference from the SBSs. Therefore, when UE~$i$ is associated with the MBS, the signal to noise ratio (SNR) at UE~$i$ will be: \vspace{-0mm}
\vspace{-2mm}
\beq\label{eq:snr}
  \gamma_{i0} ({\boldsymbol{\sigma}(t)}) =\frac{P^{\textrm{tx}}_0 h_{i0} \sigma_0(t)}{\rho^2}, 
\eeq \vspace{-4mm}\\
\noindent where $h_{i0}$ is the channel gain between UE $i$ and the MBS, and $P^{\textrm{tx}}_0$ is the transmit power of the MBS. The channel gain $h_{ij}$ can be seen as the {time-averaged} gain. 

When $\gamma_{ij}({\boldsymbol{\sigma}(t)})$ is given, UE $i$ is associated with the BS $j^*(i,{\boldsymbol{\sigma}(t)})$ that provides the largest SINR or SNR depending on whether $j^*(i,{\boldsymbol{\sigma}(t)})$ is an SBS or  MBS, respectively. Therefore, the user association can be given by: \vspace{-3mm}
\beq\label{eq:ua}
j^*(i,{\boldsymbol{\sigma}(t)})=\text{argmax}_{j \in \cB^{\textrm{on}}(t)} \gamma_{ij}({\boldsymbol{\sigma}(t)}).  
\eeq

By using the user association rule in \eqref{eq:ua}, the user association of whole network is updated at each time~$t$. 
Then, the set of UEs associated with the same BS $j$ can be defined by \vspace{-2mm}
\beq\label{eq:setue}
 \cI_j({\boldsymbol{\sigma}(t)})=\{i \;|\; j^*(i,{\boldsymbol{\sigma}(t)})=j, \forall i\}.  
\eeq \vspace{-6mm}\\
The set $\cI_j({\boldsymbol{\sigma}(t)})$ changes over time~$t$ according to the user association results from \eqref{eq:ua}. If $j\neq0$, then $\cI_j({\boldsymbol{\sigma}(t)})$ indicates the set of UEs associated with SBS $j$.  Otherwise, when $j=0$, then $\cI_0({\boldsymbol{\sigma}(0)})$ indicates the set of UEs connected to the MBS. Subsequently, the set of all UEs $\cI$ can be divided into $J+1$ subsets at most, each of which is denoted by $\cI_j({\boldsymbol{\sigma}(t)})$, $j\in\cJ$. Thus, each UE should be associated with one of BSs at any time $0 \leq t \leq T$ from \eqref{eq:ua}, and, thus, we have  $\cI =  \cup_{j=0}^{J} \cI_j({\boldsymbol{\sigma}(t)}), 0\leq t \leq T$.  

When the user association is determined by \eqref{eq:ua}, the achievable data rate of UE $i$ is given by
\beq\label{eq:datarate}
c_{ij}({\boldsymbol{\sigma}(t)})=\frac{B}{|\cI_j(t)|} \log_2(1+\gamma_{ij}({\boldsymbol{\sigma}(t)})),
\eeq 
where $|\cI_j(t)|$ is the number of UEs associated with SBS $j$ at time $t$, and $B$ is the  bandwidth of an SBS ($B=B_s$) or MBS ($B=B_m$). When the MBS can transmit data to UEs  using  bandwidth $B_m$,  time slots are scheduled for the $|\cI_0(t)|$ UEs  using a round robin scheduling. In the considered model, whenever a file of $K$ bits needs to be transmitted to each UE, we can define the total transmission delay between BS $j$ and all UE in $\cI_j(t)$ at time $t$ as  \vspace{-2mm}
\beq
\phi_j({\boldsymbol{\sigma}(t)}) = \sum_{i\in \cI_j({\boldsymbol{\sigma}(t)})}\frac{K}{c_{ij}({\boldsymbol{\sigma}(t)})}.  
\eeq\vspace{-9mm}

\subsection{Power Consumption}

Next, we define  the  power consumption models for the MBS and SBSs. When modeling the power consumption of BSs, the resource utilization of a BS   monotonically increases as the number of UE connections increases. Thus, the power consumption of a BS increase as the utilization become higher.  The power consumption model for a BS includes two components: the utilization-proportional power consumption and the fixed power consumption. The utilization-proportional power consumption depends on the signal processing functions and, hence, it varies depending on the number of associated UEs at a BS. Meanwhile,  the fixed power components pertain to the power consumed due to components such as the power amplifier or the cooler.  Thus, a fixed amount of power is required to operate the BS regardless of the number of the associated UEs. The power consumption of a BS  at time $t$ is therefore given by:  \vspace{-1mm}
\beq \label{eq:psi}
\psi_j ({\boldsymbol{\sigma}(t)}) = \frac{|\cI_j({\boldsymbol{\sigma}(t)})|}{M} (1-q) P^{\textrm{op}}_j + q P^{\textrm{op}}_j, 
\eeq
where $q$ is a weighting parameter that captures the tradeoff between the utilization-proportional power consumption and the fixed power, $P^{\textrm{op}}_j$ is the maximum power consumption when the BS is fully utilized, and $M$ is the maximum number of UE connections.  If the type of BS $j$ is a MBS, then we set $M=M_m$, and, if BS $j$ indicates an SBS, then $M=M_s$. The MBS can provide service to the larger number of UEs since the MBS has higher computing capability than an SBS; thus, the different service capabilities can be presented by $M_m \geq M_s$. Also, $P^{\textrm{tx}}_j = a P^{\textrm{op}}_j$ where the constant $a$ denotes the fraction of the transmit power $P^{\textrm{tx}}_j$ out of the total the maximum operational power $P^{\textrm{op}}_j$.  For example, if $q=1$, the BS consumes constant power regardless of the utilization level of the BS.  On the other hand, if $q=0$, the power consumption of the BS is proportional to the utilization, which is a more realistic BS power consumption model. Note that $\psi_j({\boldsymbol{\sigma}(t)})$ is the  power required to turn ON SBS $j$ at time $t$, and it depends on the number of UEs associated  with SBS $j$. 

As mentioned, SBSs use energy harvesting as a primary energy source, so an ESS can be used to store the excess  energy for future use.   The available amount of energy at time $t$ is given by \vspace{-3mm}
\be \label{eq:ehconstraint}
 \!\!\!\! E_j(t) = \min \left( \int_0^{t-\epsilon}  \Omega_j(\tau) d\tau  - \int_0^t  \psi_j({\boldsymbol{\sigma}(\tau)}) d\tau, \; E_{\textrm{max}} \right), \forall j\!\! \in\!\! \cJ,
\ee
where $E_j(t) \geq 0$ is the stored energy of SBS $j$ at time $t$, $\psi_j({\boldsymbol{\sigma}(t)})$ is the consumed power of SBS~$j$, $\Omega_j(t)$ is the amount of energy arrival of SBS~$j$, $\epsilon$ is a small number, and $E_{\textrm{max}}$ is the maximum capacity of ESS. $\Omega_j(t)$ captures the uncertainty of energy harvesting in the time domain. Since an SBS solely relies on the energy harvesting, if $E_j(t)$ becomes zero at a certain time~$t$, SBS~$j$ is turned OFF at time $t$, and the UEs connected to SBS~$j$ are handed over to other SBSs or the MBS according to the user association rule \eqref{eq:ua}.   \vspace{-5mm}

\subsection{Operational  Expenditure of Base Stations}

Given the defined network delay and power consumption models, we define operational costs incurred when using an SBS or MBS. First, we account for the operational cost of a given SBS per unit time when an SBS is turned ON. In the ON state, UEs associated with SBS $j$ experience the network delay given by $\phi_j({\boldsymbol{\sigma}(t)})$. Since higher delay is an unfavorable aspect, the operational cost has to increase with the network delay of UEs. Moreover, while an SBS is turned ON, it will incur a power consumption cost. Thus, to turn SBS $j$ ON at time $t$, the required cost of using SBS $j$ can be defined by\vspace{-2mm}
\beq\label{rental}
r_j ({\boldsymbol{\sigma}(t)})= \alpha_D \phi_j({\boldsymbol{\sigma}(t)}) + \alpha_P \psi_j({\boldsymbol{\sigma}(t)}), 
\eeq \vspace{-5mm}\\
where the constant $\alpha_D$ is the monetary cost per unit transmission delay, and the constant $\alpha_P$ is the monetary cost per unit power consumption. $\alpha_D$ and $\alpha_P$ can be used to change the weighting of delay and power consumption. The delay and energy are combined in \eqref{rental} so as to balance the tradeoff between the two metrics. The cost $r_j({\boldsymbol{\sigma}(t)})$ of a given SBS $j$ can vary over time due to the fact that the user association of UEs can change between two different times $t$ and $t'$, i.e., $\cI_j({\boldsymbol{\sigma}(t)}) \neq \cI_j({\boldsymbol{\sigma}(t')})$.\!\! Thus,  different user associations can result in  different $\phi_j({\boldsymbol{\sigma}(t)})$ and $\psi_j({\boldsymbol{\sigma}(t)})$ since the data rate of each UE and the number of connected UEs per SBS are~different. 

Next, we model the cost for using the MBS. When self-powered SBSs rely solely on the harvested energy that is highly uncertain and intermittent, they might need to turn OFF if they have no more energy. Therefore, to avoid the risk of such energy depletion, the SBSs can go into an energy-saving OFF state to store additional energy for future use. Due to this energy storage need, the system can end up with a large number of OFF SBSs which, in turn, will degrade the network performance as it increases congestion at the MBS and the ON SBSs. Thus, to prevent such a network congestion, if SBS $j$ decides to switch OFF, we assume that it will be charged a cost $b_j$. By setting a flat-rate cost $b_j$, the network can control how often the SBSs can turn OFF, particularly when they still have a sufficient amount of energy stored. Here, as $b_j$ increases, the penalty of turning a given SBS $j$ OFF becomes larger; thus, the SBSs will have an incentive to maintain the ON state as long as possible. In a dynamic network, the ON and OFF states of the SBSs can change over time thus also changing the user association. In such a dynamic network, finding an exact, flat rate $b_j$ is difficult. Therefore, we propose to derive this cost based on a worst-case assumption. In particular, to define the cost $b_j$, first we find the maximum cost of using the MBS which is then scaled by a parameter $\alpha_B \in [0,1]$. The cost $b_j$ is the maximum cost that can be incurred by turning OFF and transferring  traffic to the MBS. To find the maximum cost of using the MBS in the worst case, suppose that all UEs can be associated with the MBS so that the network delay and power consumption of the MBS are maximized. Here, when a portion of the maximum cost is incurred to an SBS, the incurred cost can depend on the UEs in the SBS denoted by the set $\cI_j({\boldsymbol{\sigma}(0)})$. By doing so, the maximum cost of using the MBS can be divided into the per-SBS costs. If  UE $i \in \cI_j(0)$ is connected to the MBS, the transmission delay of UE $i$ will be $\frac{K}{\frac{B_m}{I} \log_2 (1+\gamma_{i0}(0)) }$.  By summing over all UEs in $\cI_j(0)$, we  obtain the network delay corresponding to the UEs in $\cI_j(0)$, as shown as \vspace{-3mm}
\beq
\Phi^{\cI_j(0)}_0 = \sum_{i \in \cI_j(0)}\frac{K}{\frac{B_m}{I} \log_2 (1+\gamma_{i0}(0)) }.
\eeq
Also, the portion of the power consumption of the MBS that is consumed by the UEs in $\cI_j(0)$ will be:\vspace{-2mm}
\beq
\Psi^{\cI_j({\boldsymbol{\sigma}(0)})}_0 = \frac{|\cI_j({\boldsymbol{\sigma}(0)})|}{M}(1-q) P^{\textrm{op}}_0 + q P^{\textrm{op}}_0.
\eeq
Consequently, whenever an SBS $j$ decides to turn OFF, the accompanying cost, due to the handover to the MBS, will be given by:\vspace{-3mm}
\beq\label{buy}
b_j=\alpha_B \left(\alpha_D \Phi^{\cI_j({\boldsymbol{\sigma}(0)})}_0  + \alpha_P  \Psi^{\cI_j({\boldsymbol{\sigma}(0)})}_0  \right)T,
\eeq \vspace{-5mm}\\
where $\alpha_B \in [0,1]$ is the fraction of the maximum cost. For example, when we set $\alpha_B=0.10$, then $10\%$ of the maximum cost of using the MBS during time period $T$ will be incurred to SBS~$j$. Thus, if the value of $b_j$ is too high, being turned ON becomes an affordable option, so  SBS~$j$ is turned ON until the whole harvested energy is used. On the other hand, if the value of $b_j$ is low,  SBSs tend to be turned OFF to keep the stored harvested energy due to a low penalty in switching SBSs OFF.  \vspace{-5mm}

\section{Problem Formulation}\label{sec:problemformulation}

Given the operational costs, our goal is to analyze the optimal ON and OFF scheduling problem for the SBSs. In cellular networks consisting of self-powered SBSs, the amount of available energy is dynamically changing and very limited. To be able to operate using energy harvesting as a primary energy source of SBSs, self-powered SBSs should intelligently manage their ON and OFF states considering delay, power, and energy state. Moreover, since future energy arrivals can be highly unpredictable, optimizing the ON and OFF schedule of SBSs is a very challenging problem. By properly scheduling its OFF duration, an SBS can reduce its  energy consumption while also storing more energy for future use. However, at the same time, the SBS must turn ON for a sufficient period of time to service users and offload MBS traffic. In our problem, information on energy arrival is unknown, so an online optimization approach is suitable. To cope with the inherent uncertainty of energy harvesting while balancing the tradeoff between energy consumption and network delay, we introduce a novel, self-organizing \emph{online optimization framework} for optimizing the ON and OFF schedule of  self-powered SBSs. \vspace{-3mm}

\subsection{ON/OFF Scheduling as an Online Optimization Problem}

We formulate the global ON and OFF scheduling problem with the goal of minimizing the sum of costs that encompass the costs of using an SBS and the MBS in \eqref{rental} and \eqref{buy}, as follows: \vspace{-3mm}
\beq \label{problem1}
  \min_{\boldsymbol{\sigma}(t), \boldsymbol{x}}&&  \sum_{j=1}^J \left( \int_{0}^{u_j} r_j({\boldsymbol{\sigma}(\tau)}) \sigma_j(\tau)d\tau + b_j x_j  \right),\\
  \textrm{s.t.}&& \sigma_j(t) + x_j \geq 1,  \;\;0 \leq t \leq u_j,\; \forall j,\label{problem1_c1}\\
                    && \sigma_j(t) \in \{0, 1\}, \;\; 0 \leq t \leq u_j,\; \forall j,\label{problem1_c2}\\
                    && x_j \in \{0,1\}, \;\;  \forall j,\label{problem1_c3}
\eeq \vspace{-5mm}\\
where ${\boldsymbol{x}} = [x_j|\forall j \in \cJ]$, respectively. The ON and OFF states of SBS $j$ at time $t$ is denoted by $\sigma_j(t)$ in \eqref{problem1_c2}. Also, $x_j$ in \eqref{problem1_c3} indicates whether SBS $j$ is determined to be turned OFF before SBS $j$'s stored energy is depleted at time~$u_j$. In \eqref{problem1_c1} and  \eqref{problem1_c2}, time $t>u_j$ is not considered since SBS $j$ is turned OFF due to  energy depletion. Note that $u_j$ is the first moment when energy harvesting constraint  \eqref{eq:ehconstraint} is not satisfied. Thus, each SBS can experience energy depletion at a different time $u_j$ since the amount of energy arrival of SBS $j$ denoted by $\Omega_j(t)$ is unknown before time $t$, and SBS $j$ cannot know the future energy status, as observed in many real-world scenarios \cite{chen2014rechargeable}. For example, when energy is harvested from the  environment, the amount of harvested energy can quickly change due to factors such as  weather conditions which can change rapidly during are changing in a short period of time. Not only the sudden weather, long-term seasonal changes also brings uncertainty into energy harvesting. Therefore,  the uncertainty of the harvested energy at each moment can be captured by $\Omega_j(t)$, and, thus, the energy depletion time $u_j$ is \emph{unknown} in our problem. In essence,  our problem is online where energy harvesting brings in uncertainty about the future event. The period $T$ can be defined in various ways.  For example, $T$ can be defined as a short period of time during which the SBS can stay ON using a fully charged battery. 

Also, it is required to reduce the network congestion by increasing the use of the harvested energy, so the ON time of each SBS needs to be extended. In problem \eqref{problem1}, if an SBS is turned OFF due to energy depletion, the cost of using the MBS is not incurred to the SBS so as to provide incentives for SBSs to maintain a longer ON period. However, if SBS $j$ is turned OFF according to its decision, the cost of using the MBS is incurred to the SBS, as captured by setting $x_j=1$. Therefore, the ON and OFF scheduling solution given by $\sigma_j(t)$ and $x_j$ can be determined by SBS $j$ during $1 \leq t \leq u_j$ so that UEs in {\color{black}$\cI_j({\boldsymbol{\sigma}(t)})$} can be connected to either SBS $j$ ($\sigma_j(t)=1$) or the MBS  ($x_j=1$) by satisfying  constraint \eqref{problem1_c1}. 

If the problem is offline, then it can be readily solved. For example, in the offline scenario, the optimal solution is either always ON strategy ($\sigma_j(t)=1, 0\leq t \leq u_j, x_j=0$) or OFF strategy ($\sigma_j(t)=0, 0\leq t \leq u_j,  x_j=1$). When $u_j$ is known in offline, it is possible to compute the total costs corresponding to a strategy that the SBS uses. 
Thus, since the SBS can compare the costs of all possible solutions, the optimal solution can be found. However, such offline scenario is not available in real environment due to the uncertainty of energy harvesting as mentioned above. 
Thus, the problem \eqref{problem1} needs to be considered in an online optimization framework. 

To solve \eqref{problem1}, one must develop a suitable online algorithm. To assess the effectiveness of such an algorithm, we need to use competitive analysis. Competitive analysis \cite{grotschel2013online} is a method used to compare between the performance of online algorithms and that of an optimal offline algorithm. One key metric in competitive analysis is the so-called competitive ratio, defined next:\vspace{-2mm}
\begin{definition}\normalfont\label{eq:cr}
The \emph{competitive ratio} of an online algorithm  is defined by \vspace{-3mm}
\beq \label{eq:defcr}
  \kappa = \max_{u_j}  \frac{\beta_{\textrm{ALG}}(u_j)}{\beta_{\textrm{OPT}}(u_j)}, \;\; \forall u_j, 
\eeq\vspace{-4mm}\\
where $u_j$ is a random time instant when harvested energy is depleted, $\beta_{\textrm{ALG}} (u_j)$ is the cost of an online algorithm that corresponds to the total cost of the problem \eqref{problem1}, and $\beta_{\textrm{OPT}}(u_j)$ is the optimal cost achieved by using an offline algorithm that knows all input information.  
\end{definition}

When we use an online algorithm, our goal is to find an algorithm that minimize the competitive ratio $\kappa$. Therefore, in competitive analysis, the competitive ratio is meaningful since it shows the performance of an online algorithm \cite{borodin2005online}. For this analysis, the competitive ratio of online algorithms is evaluated for a given arbitrary input sequence that corresponds to uncertain energy arrivals. In our model, the arbitrary input sequence is characterized by $u_j$ that is the moment of energy depletion. From the competitive analysis, even though an SBS does not know the input sequence, the use of online algorithms will give a solution that can at least achieve the cost of $ \kappa  \beta_{\textrm{OPT}}(u_j)$. 

To analyze this problem, first, we consider two  special cases in which: a) ${\color{black}r_j({\boldsymbol{\sigma}(t)})}$ is decreasing over time or b) ${\color{black}r_j({\boldsymbol{\sigma}(t)})}$ is increasing over time. If an SBS's ${\color{black}r_j({\boldsymbol{\sigma}(t)})}$ decreases, the SBS can have  motivation to extend its ON time since the cost of using SBS becomes inexpensive. Thus, the SBS can simply extend the ON time. On the other hand, if ${\color{black}r_j({\boldsymbol{\sigma}(t)})}$ increases, the SBS has less motivation of maintaining the ON state. Moreover, in this case, it is possible that the SBS could stay in the OFF state from the beginning if the SBS knew the increasing of  ${\color{black}r_j({\boldsymbol{\sigma}(t)})}$. Therefore, since the SBS cannot change its previous decisions in the case in which ${\color{black}r_j({\boldsymbol{\sigma}(t)})}$ is  increasing, it is difficult to minimize the total cost. 

Thus, we present an example case where the cost of using an SBS ${\color{black}r_j({\boldsymbol{\sigma}(t)})}$ decreases as the time~$t$ increases. By doing so, we can propose an ON and OFF scheduling algorithm that achieves a finite competitive ratio. Note that the decreasing of ${\color{black}r_j({\boldsymbol{\sigma}(t)})}$ can be physically observed when an SBS increases the transmission power so that it can decrease the delay cost of the SBS as shown in our simulations. In such case, we propose an online algorithm in which the SBS is turned OFF at a predetermined time $\bar{t}$. When the value of ${\color{black}r_j({\boldsymbol{\sigma}(t)})}$ decreases, each achieved value for  ${\color{black}r_j({\boldsymbol{\sigma}(t)})}$ will be denoted by $r_{(v)}$. These values are then arranged in a descending order where $v$ indicates the order of a given value $r_{(v)}$, as follows: \vspace{-2mm}
\beq\label{eq:decreasing_r}
  r_{(1)} >   r_{(2)} >  \cdots > r_{(v-1)} >   r_{(v)}. 
\eeq\vspace{-5mm}\\
Here, we note that, ${\color{black}r_j({\boldsymbol{\sigma}(t)})}$ changes from $r_{(v-1)}$ to $r_{(v)}$ at time $t_{(v-1)}$, and $r_{(v)}$ stays constant from $t_{(v-1)}$ to $t_{(v)}$ where $t_{(0)}=0 <  t_{(1)} <   t_{(2)} <  \cdots < t_{(v-1)} < t_{(v)}$. 

\vspace{-2mm}
\begin{theorem}\label{theorem1}
When ${\color{black}r_j({\boldsymbol{\sigma}(t)})}$ decreases over time $t$ in the problem \eqref{problem1},  the initial SBS's OFF time is given by $\bar{t}={b_j}/{r_{(1)}}$ at time $t_{(0)}$. Also, at time $t_{(v-1)}$, $v \geq 2$, the SBS's OFF time is updated using the following equation:\vspace{-2mm}
\beq\label{eq:bart}
  \bar{t}=\frac{b_j}{r_{(v)}} -  \frac{1}{r_{(v)}} \sum_{v'=1}^{v-1} t_{(v')} \left(r_{(v')}-r_{(v'+1)}\right). 
\eeq\vspace{-3mm}\\
Then, the OFF time $\bar{t}$ increases when it is updated by \eqref{eq:bart}. Also, an online OFF time scheduling algorithm that uses $\bar{t}$ can achieve a competitive ratio of $2$. 
\end{theorem}\vspace{-5mm}
\begin{proof}See the Appendix. 
\end{proof}
In Theorem~\ref{theorem1}, at the time in which the SBS's cost ${\color{black}r_j({\boldsymbol{\sigma}(t)})}$ is updated, the SBS  update its ON time by setting a larger value for  $\bar{t}$. Thus, the updated $\bar{t}$ effectively optimizes the problem. 

To investigate more  dynamically changing ${\color{black}r_j({\boldsymbol{\sigma}(t)})}$ needs to be considered. However, since the value of ${\color{black}r_j({\boldsymbol{\sigma}(t)})}$ depends on the ON/OFF state of SBSs in a network, the exact value of a future ${\color{black}r_j({\boldsymbol{\sigma}(t)})}$ cannot be known and expected. For instance, if the neighboring SBSs are turned OFF, the interference at SBS $j$ will be reduced thus increasing the data rate of UEs that are associated with SBS $j$. This, in turn, results in a smaller delay cost and reduces ${\color{black}r_j({\boldsymbol{\sigma}(t)})}$. At the same time, UEs associated with other,  neighboring SBSs may be handed over to SBS $j$. Then, the number of UEs served by SBS $j$ increases thus increasing the delay cost.  In addition, due to the increase of the number of UEs, the power consumption of SBS~$j$ also increases thus yielding a higher ${\color{black}r_j({\boldsymbol{\sigma}(t)})}$. As seen from these illustrative scenarios, the OFF scheduling of the various SBSs can either increase or decrease ${\color{black}r_j({\boldsymbol{\sigma}(t)})}$. Therefore, the cost of using a given SBS will not always be monotonically increasing or decreasing thus making it very challenging to find a solution to the optimization problem in \eqref{problem1} by estimating the future variation of ${\color{black}r_j({\boldsymbol{\sigma}(t)})}$ over time $t$. Moreover, to solve \eqref{problem1}, the ON and OFF states of all SBS must be collected by the network which  can generate additional signaling overhead for information exchange. This can also require the use of a centralized controller. Naturally, in a dense SCN, such centralized control may not be possible or scalable. 

Consequently, in essence, our goal is to devise a self-organizing approach in which the solution to \eqref{problem1} can be done locally at each SBS. Clearly, solving this problem for a generic, non-monotonically changing ${\color{black}r_j({\boldsymbol{\sigma}(t)})}$ is challenging and, therefore, we need to use an approximation. One natural way is to assume that ${\color{black}r_j({\boldsymbol{\sigma}(t)})}$ is not time-varying, which can simplify the problem because the interference and user association that change over time do not need to be considered, as discussed next. \vspace{-5mm}

\subsection{Approximated Problem}

To relax the time dependence from $r_j({\boldsymbol{\sigma}(t)})$, we assume that the cost will be equal to $r_j=r_j({\boldsymbol{\sigma}(0)})$. In other words, the initial cost, which is generally known to the network, will be used as a flat cost of using an SBS. This approximation can help simplify the problem by considering a worst-case assumption for the interference, as follows. As mentioned, the cost $r_j({\boldsymbol{\sigma}(t)})$ incurred to an SBS $j$ is affected by interference when other SBSs are randomly turned OFF. However, by approximating $r_j({\boldsymbol{\sigma}(t)})$ using a constant value, the scheduling decisions will no longer be dependent and, thus, each SBS can make its own decision without having global knowledge about other SBSs' ON and OFF states. Note that the largest value of the interference is captured in the approximated problem since all SBSs are turned  ON at the beginning. 
Thus, the  SBSs can compute the value of $r_j$  even though all SBSs are not actually turned ON. One key advantage of the proposed approach is that an SBS can determine the solution at the beginning of each period~$T$. Thus, distributed optimization can be done by computing locally, and also it reduce network overhead since signaling is not required. Here, the approximated problem can be given by:\vspace{-4mm}
\beq \label{problem2}
  \min_{\boldsymbol{\sigma}(t), \boldsymbol{x}}&&  \sum_{j=1}^J \left( \int_{0}^{u_j}r_j \sigma_j(\tau)d\tau + b_j x_j  \right),\label{}\\
 \textrm{s.t.} &&  \eqref{problem1_c1}, \eqref{problem1_c2}, \textrm{and} \; \eqref{problem1_c3}.\nonumber
\eeq \vspace{-5mm}

To solve problem \eqref{problem2}, we decompose it  into smaller, per SBS subproblems. As shown next, each SBS can solve an individual optimization subproblem, so the approximated problem in \eqref{problem2} can be solved in a distributed way. \vspace{-2mm}
\begin{proposition} 
The problem in \eqref{problem2} can be decomposed into $|\cJ|$ subproblems.
\end{proposition}\vspace{-5mm}
\begin{proof}
The objective function of the problem \eqref{problem2} can be shown to be a sum of functions of $\sigma_j(t)$ and $x_j$ as shown as $\eqref{problem2}$. Thus, changing of $\sigma_j(t)$ and $x_j$ does not affect $\sigma_{j'}(t)$ and $x_{j'}$, $j' \neq j$. Therefore, the objective function of \eqref{problem2} can be separated into $|\cJ|$ functions. Also, each SBS's energy storage is not connected to other SBSs' energy source. Thus, due to the isolated energy harvesting system of each SBS, the amount of stored energy shown as \eqref{eq:ehconstraint} is managed independently by each SBS. Hence, the problem \eqref{problem2} can be decomposed into $|\cJ|$ subproblems. 
\end{proof}
Now, we have $|\cJ|$ subproblems derived from the approximated problem in \eqref{problem2}. The ON or OFF decision of an SBS does not affect the decision of another SBS, so  we can solve  $|\cJ|$ subproblems in parallel. By solving each of the per-SBS problems, we can significantly reduce complexity and overhead while allowing for a self-organizing implementation. Consequently, each SBS will solve its local version of \eqref{problem2} that seeks to minimize its individual cost function given by \vspace{-2mm}
\beq\label{problem3}
 \min_{\sigma_j(t),x_j}&&  \int_{0}^{u_j} r_j \sigma_j(\tau)d\tau+ b_j x_j,\\
 \textrm{s.t.} && \eqref{problem1_c1}, \eqref{problem1_c2}, \textrm{and} \; \eqref{problem1_c3}.\nonumber 
\eeq \vspace{-6mm} \\
Since SBS $j$ does not know the whole input sequence (e.g., uncertain energy arrivals), the SBS cannot know the optimal schedule of ON and OFF before time elapses. Thus,  \eqref{problem3} is still formulated as an online optimization problem, for which an online algorithm is needed to make a decision in real time under an uncertain future. Remarkably, the problem in \eqref{problem3} is analogous to the so-called \emph{ski rental problem} \cite{lotker2008ski}, an online optimization framework that enables such decision making in face of uncertainty, as discussed next. 

\vspace{-3mm}
\section{On/Off Scheduling as an Online Ski Rental Problem}\label{sec:skirental}

First, we will explicitly define the analogy between ski rental and self-powered BS scheduling. In the classical online ski rental problem, an individual is going skiing for an unknown number of days \cite{grotschel2013online}. The uncertainty on the skiing period is due to factors such as nature or whether this individual will enjoy skiing or not. Here, the individual must decide on whether to rent skis over a short period of time or, alternatively, buy them for a long period of time, depending on the costs of renting and buying, the number of days that he/she will end up skiing, and on whether the skiing activity will be enjoyable. The online ski rental framework provides online optimization techniques that allows one to understand how an individual will make a ``rent'' or ``buy'' decision in such a scenario while facing  uncertainty due to nature and while accounting for the tradeoff between the costs of rental and purchase  and the benefits of skiing. 
 
In this regard, our problem in \eqref{problem3} is similar to the ski rental decision making process. In our model, each SBS is an individual that must   \emph{rent}  its resources (turn ON) to the network under the uncertainty of energy harvesting or alternatively  \emph{buy}  more reliable MBS resources (and turn OFF). From \eqref{rental} and \eqref{buy}, we can see that  $r_j$ and $b_j$  will represent the prices for rent and buy,  respectively. Thus, the decision of an SBS on how long to  turn ON is essentially a decision on how long to rent its resources which require paying $r_j$ per unit time.      
Once the SBS turns OFF, the network must buy  the more expensive but more reliable MBS resources at a price $b_j$. Given this analogy, we can develop efficient online algorithms to solve  \eqref{problem2} \cite{albers2010energy}. An {online algorithm} can solve the problem at each present time without having whole information about future energy harvesting results.      

To solve the BS ON/OFF scheduling problem, one may consider other methods such as  Markov decision processes, dynamic programming, reinforcement learning, or convex online optimization. However, those are not suitable frameworks for studying the problem considered in this work since additional assumption or information on energy harvesting process would be required to model the environment.  

We use online algorithms to solve the optimization problem, and competitive analysis is used to study the performance of the online algorithms. We first analyze the optimal offline strategy when assuming  energy arrival information over the entire  period is given. The offline optimal cost can be shown as\vspace{-4mm}
\beq\label{eq:offlinecost}
			\beta_{\textrm{OPT}} (u_j)= ~\left\{\begin{matrix}
			r_j u_j,& 0 \leq u_j \leq \frac{b_j}{r_j}, \\ 
			b_j,& \frac{b_j}{r_j} \leq u_j \leq T.  \\
			\end{matrix}\right.
\eeq \vspace{-5mm}\\
The optimal solution is using the rent option until ${b_j}/{r_j}$ if energy is depleted earlier than ${b_j}/{r_j}$. Otherwise, the buy option should be chosen with one time payment $b_j$ at time $0$.  

\vspace{-4mm}
\subsection{Deterministic Online Algorithm}

\begin{figure}[t]
\centering
\includegraphics[width=7cm]{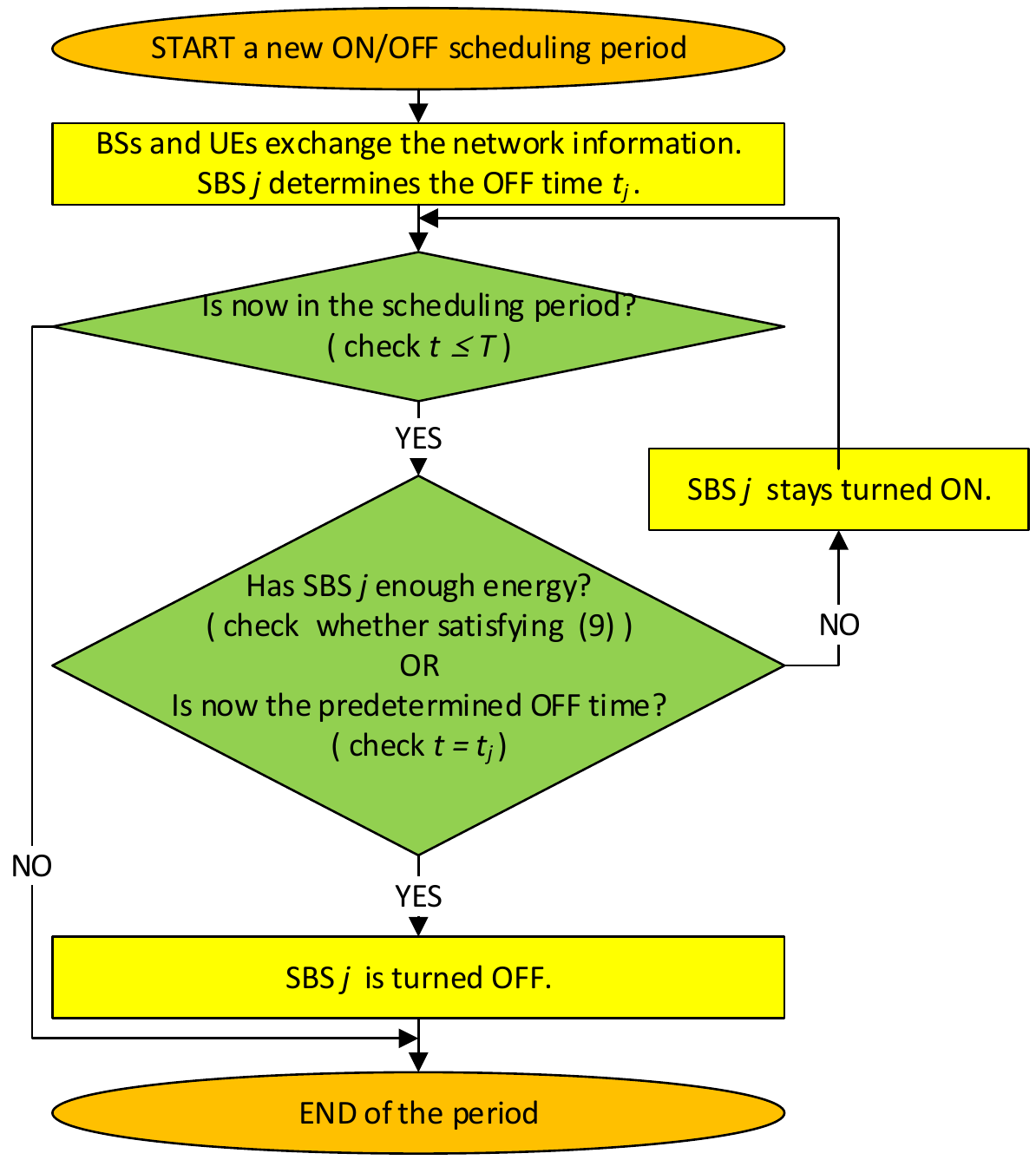}\vspace{-2mm}
\caption{\small  {\color{black}Flowchart of Algorithm~1.}}
\label{fig:flowchart} \vspace{2mm}
\end{figure}

To design an online algorithm that can achieve a close performance  to  optimal, we first investigate how close performance a deterministic online algorithm can yield. A deterministic approach is mainly operated by a predetermined parameter when making decision of ON/OFF scheduling.   
\begin{algorithm}[t]
\caption{Deterministic Online Algorithm (DOA)}\label{algorithm:baseline}
\begin{algorithmic}[1]
\item[1:] Initialization: SBS $j\in\cJ$ has a predetermined value $t_j=b_j/r_j$.
\item[2:] {\bf while} $t \leq T$ 
\item[3:] \hspace{0.3cm} Update $t\leftarrow t+\epsilon$.
\item[4:] \hspace{0.3cm} {\bf If} (\eqref{eq:ehconstraint} is unsatisfied) or ($t = t_j$),
\item[5:] \hspace{0.6cm}   {\bf then} SBS $j$ is turned OFF. 
\item[6:] \hspace{0.3cm} {\bf else} SBS $j$ maintains its ON state.
\item[7:] {\bf end while}
\item[8:] At $t=T$, update $P^{\textrm{op}}_j, P^{\textrm{tx}}_j, \forall j\in\cJ$, and user association.
\end{algorithmic}
\end{algorithm}\smallskip
In a deterministic online algorithm (DOA), SBS $j$ is turned OFF at a predetermined time $t_j$,  $0 \leq t_j \leq T$.  
This flowchart in Fig.~\ref{fig:flowchart} shows the structure of  Algorithm~1 where the OFF time is determined at the beginning of the period. From time $0$ to $t_j$, the rent option is used, and the cost increases along with the rental cost $r_j$ per time. Then, at time $t_j$, the buy option is purchased for the one time cost $b_j$. DOA  can be shown as Algorithm~\ref{algorithm:baseline}. The competitive ratio $  \kappa $ of DOA  is given by\vspace{-2mm}
\beq\label{eq:optkappa}
  \frac{\beta_{\textrm{DOA}}(u_j)}{\beta_{\textrm{OPT}}(u_j)} 
= ~\left\{\begin{matrix}
	\frac{r_j u_j}{\min\{r_j u_j, b_j\}} ,& 0 \leq u_j \leq t_j, \\ 
	\frac{r_j t_j + b_j}{\min\{r_j u_j, b_j\}},& t_j \leq u_j \leq T,  \\
	\end{matrix}\right. 
\eeq\vspace{-3mm}\\
where $\beta_{\textrm{DOA}}$ is the cost of DOA. 

We want to minimize $\kappa$ subject to ${\beta_{\textrm{DOA}}(u_j)} \leq \kappa {\beta_{\textrm{OPT}}(u_j)} $ for every $u_j$ from $0$ to $T$. Therefore, when $u_j = t_j = b_j/r_j$, the competitive ratio becomes 2  known as the best possible competitive ratio of a deterministic, online algorithm \cite{lotker2008ski}. 

\vspace{-4mm}
\subsection{Randomized Online Algorithm} \label{sec:online_algorithm}

To handle  uncertainty, a rent or buy decision will be made by using a randomized online algorithm (ROA) by means of  a probability distribution for ON/OFF scheduling  designed to solve our cost-minimization problem. For instance, it is known that, when a randomized approach is used to address a ski rental problem,  it is possible to achieve a lower competitive ratio of $\frac{e}{e-1}$ \cite{karlin1994competitive, lotker2008ski}, while DOA achieves the competitive ratio of 2.

To develop an ROA  for our problem,  a competitive analysis analogous to the one done in \cite{lotker2008ski} will be followed. For an arbitrary input,  ROA computes an output (i.e., the turn OFF time, $t_j$) based on a probability distribution. We want to design an ROA that satisfies $\E[F_j(t_j)] < \kappa  \beta_{\textrm{OPT}}(u_j) $ where $\E[F_j(t_j)]$ is the expected cost of the problem \eqref{problem3} redefined by $ F_j(t_j) = \big\{ 
													\begin{array}{ll}
          						     r_j u_j, &\textrm{if } u_j < t_j, \vspace{-2mm}\\
						               r_j t_j + b_j, &\textrm{if } u_j \geq t_j,
			                \end{array} $ provided that unknown time of energy depletion is given by $u_j$. This will be adequate for our problem in that the input sequence is the unknown and uncertain energy arrivals at a given SBS. 
Even though an SBS does not know the input sequence, the use of an ROA will give a solution that can at least achieve the expected cost of $ \kappa  \beta_{\textrm{OPT}} $.

In this section, when the rental price $r_j$ and the buying price $b_j$ are values related to the cost of using an SBS and the MBS, respectively, we will compute the expected cost of  ROA. At time $t_j$, the state of the SBS can be either ON or OFF with  probability distribution $p^{\textrm{on}}_j(t_j)$ or $p^{\textrm{off}}_j(t_j) = 1 - p^{\textrm{on}}_j(t_j)$.     
When an SBS  decides to turn OFF at $t_j$,  we have \vspace{-3mm}
\be\label{eq:expectedcost}
   \E[F_j(t_j)] =\! \int_0^{u_j} \!(r_j t_j + b_j)  p'^{\textrm{off}}_j(t_j) d t_j \!+\!  \int_{u_j}^T \! r_j u_j p'^{\textrm{off}}_j (t_j) d t_j,
\ee \vspace{-4mm}\\ 
where $p'^{\textrm{off}}_j(t_j)$ is the first-order derivative of $p^{\textrm{off}}_j(t_j)$. Then, from $\frac{d}{d u_j} \E[F_j(t_j)] =R_j(u_j) $,  the rate of increase of the cost will be expressed by \vspace{-3mm}
\be
R_j(u_j) = r_j p^{\textrm{on}}_j(u_j)  +  r_j u_j p'^{\textrm{on}}_j(u_j) +  (r_j u_j  +  b_j) p'^{\textrm{off}}_j(u_j),\nonumber
\ee \vspace{-5mm}\\
where  $p'^{\textrm{on}}_j=-p'^{\textrm{off}}_j$. To  find an upper bound on  $F_j(t_j)$, we focus on the case in which the expected cost is at its  largest value. Naturally, this is the same as finding the worst case in the online ski rental problem which corresponds to the case in which the individual buys the skis on one day, but is unable to use them in the next day.     
In our model, this corresponds to the case in which the SBS pays for the MBS resources at a price $b_j$ at $u_j$ due to  the uncertainty of energy. However, at $u_j=t_j$, the SBS does not need to turn OFF if new energy arrives suddenly at that moment. In this worst case, the cost-increasing rate $R_j(u_j)$ becomes\vspace{-3mm}
\beq
R_j(t_j) &=&  r_j p^{\textrm{on}}_j(t_j)  +  r_j t_j p'^{\textrm{on}}_j(t_j) +  (r_j t_j  +  b_j) p'^{\textrm{off}}_j(t_j)\nonumber\\
&=&  r_j p^{\textrm{on}}_j(t_j)  -  b_j p'^{\textrm{on}}_j(t_j) \nonumber.
\eeq \vspace{-6mm}\\
By using the relationship $\E[F_j(t_j)] < \kappa  \beta_{\textrm{OPT}}$, the cost-increasing rate of $\E[F_j(t_j)] $ cannot be larger than the cost-increasing rate of $\kappa \beta_{\textrm{OPT}}$. The cost-increasing rate of $ \beta_{\textrm{OPT}}$ with respect to $u_j$ can be readily derived by choosing the rent or buy option that yields smaller cost. Now, we divide the range of $u_j,  t_j$ into two cases.   

First, if $0 < u_j < {b_j}/{r_j}$ and $0 < t_j < {b_j}/{r_j}$, then the optimal cost-increasing rate is $r_j$ which means that an SBS should be turned ON during $t_j$.   Thus, the cost-increasing rate of ROA cannot be lower than $\kappa$ times the optimal cost-increasing rate, we have 
\vspace{-3mm}
\beq \label{eq:ode}
r_j \kappa = r_j p^{\textrm{on}}_j(t_j) -  b_j p'^{\textrm{on}}_j(t_j). \nonumber
\eeq
\vspace{-6mm}\\
Since this is a first-order linear ordinary differential equation, the solution $p^{\textrm{on}}_j(t_j)$ is given by:   
\vspace{-3mm}
\beq
p^{\textrm{on}}_j(t_j) =c e^{\frac{r_j}{b_j}t_j} + \kappa,
\eeq
\vspace{-6mm}\\
where $c$ is a constant that can be found by using two boundary conditions. If an SBS starts with the ON state, then $p^{\textrm{on}}_j(0) = \kappa + c = 1$, and then $c = 1 - \kappa$. 

\begin{algorithm}[t]
\caption{Randomized Online  Algorithm (ROA)}\label{algorithm:online}
\begin{algorithmic}[1]
\item[1:] Initialization: SBS $j\in\cJ$ determines $r_j$ and $b_j$.
\item[2:] Find $t_j$ s.t. $ p^{\textrm{off}}_j(t_j) = \mu_j$, $\mu_j\!\sim\!U\!(0,1), \forall j\in\cJ$.
\item[3:] {\bf while} $t \leq T$ 
\item[4:] \hspace{0.3cm} Update $t  \leftarrow t+\epsilon $.
\item[5:] \hspace{0.3cm} {\bf If} (\eqref{eq:ehconstraint} is unsatisfied) or ($t = t_j$),
\item[6:] \hspace{0.6cm}   {\bf then} SBS $j$ is turned OFF. 
\item[7:] \hspace{0.3cm} {\bf else} SBS $j$ maintains its ON state.
\item[8:] {\bf end while}
\item[9:] At $t=T$, update $P^{\textrm{op}}_j, P^{\textrm{tx}}_j, \forall j\in\cJ$, and user association.
\end{algorithmic}
\end{algorithm}\smallskip

Second, if ${b_j}/{r_j} < u_j$ and ${b_j}/{r_j} <  t_j$, then using the MBS is the optimal choice. In this case, an SBS should buy the MBS resource  before ${b_j}/{r_j}$. Thus, the SBS should remain in the OFF state at ${b_j}/{r_j}$. This fact leads us to find $p^{\textrm{on}}_j({b_j}/{r_j}) = (1-\kappa)e+\kappa = 0$, and we find $\kappa=\frac{e}{e-1}$. Therefore, we have the ON probability $p^{\textrm{on}}_j(t_j) = \frac{ e-e^{\frac{r_j}{b_j} t_j} }{e-1}$. \vspace{-2mm} 
\begin{remark}\normalfont
At $t_j$, SBS $j$ will turn OFF according to the following probability distribution,\vspace{-1mm}
\beq\label{eq:Proff}
			p^{\textrm{off}}_j(t_j) = ~\left\{\begin{matrix}
			\frac{ e^{\frac{r_j}{b_j} t_j}-1 }{e-1},& 0 \leq t_j \leq \frac{b_j}{r_j}, \\ 
			1,& \frac{b_j}{r_j} \leq t_j \leq T.  \\
			\end{matrix}\right.
\eeq
\end{remark}\vspace{-2mm}
\noindent The proposed online ski rental algorithm is summarized in Algorithm~\ref{algorithm:online}. From \eqref{eq:Proff}, we observe the tradeoff between rent and buy. As mentioned, the rental price is a cost related to using an SBS while the buying price reflects the cost of using the MBS. For example, the rental price is reduced if using an SBS yields lower delay cost, or the power consumption of an SBS is reduced. Also, the buying price increases if the delay from using the MBS increases, or the power consumption of the MBS increases. Therefore, if $r_j$ is low and $b_j$ is high, then it implies that using SBS will reap benefits in terms of delay cost or power consumption, so  the rent time becomes longer.    
In contrast, the rent time becomes shorter if $r_j$ is high and $b_j$ is low. The short rent time means an SBS turns OFF early because buying the MBS resource would be more beneficial  than using the SBS resource with the rent price. Each SBS will now run Algorithm~\ref{algorithm:online} and  decide at time $t=0$ when to turn OFF, without knowing any information on energy arrivals, by using the distribution in \eqref{eq:Proff}. From \eqref{eq:Proff}, we can observe that the OFF time can be adjusted by changing the value of $T$.  For example, if $b_j/r_j$ increases by having a longer period of $T$, the ON time can be extended, so it can prevent the frequent ON/OFF switching. Also, it can be helpful to reduce the frequent handovers. 

When using the Algorithm~\ref{algorithm:online}, we can verify that the expected competitive ratio of ROA is $\frac{e}{e-1}$ if $r_j T \geq b_j$ is satisfied. When the rental option is chosen during the whole period $T$,  the total cost is $r_j T$. If the total cost is smaller than selecting the buy option such that $r_j T < b_j$, then this leads to a special case. For such a case, since the optimal solution is always choosing the rental option, the SBS is not turned OFF until the energy is exhausted. Therefore, to find a solution of our interest, we should consider the case in which $r_j T \geq b_j$.     

Then, to show the expected competitive ratio,  we calculate the expected cost of ROA. First, let us consider when $0 \leq u_j < b_j/r_j$ and $b_j/r_j <T$. By using \eqref{eq:expectedcost}, the expected cost is\vspace{-3mm}
\beq
&   \E[F_j(t_j)] = \int_0^{u_j} (r_j t_j + b_j)  p'^{\textrm{off}}_j(t_j) d t_j  + &\nonumber\\
&    \int_{u_j}^{\frac{b_j}{r_j}}  r_j u_j p'^{\textrm{off}}_j (t_j) d t_j +  \int_{\frac{b_j}{r_j}}^T  r_j u_j p'^{\textrm{off}}_j (t_j) d t_j   =  \frac{r_j u_j e}{e-1}, &\label{eq:expectedcost1int}
\eeq\vspace{-4mm}\\
{\noindent}where 
 $ p'^{\textrm{off}}_j(t_j) = \Big\{ 
													\begin{array}{ll}
          						     \frac{r_j}{b_j} \frac{ e^{\frac{r_j}{b_j} t_j}}{e-1},& 0 \leq t_j \leq {b_j}/{r_j}, \vspace{-0.5mm}\\
						               0,& {b_j}/{r_j} \leq t_j \leq T.\vspace{0mm}\end{array}$.
The third integration in \eqref{eq:expectedcost1int} becomes zero since $p'^{\textrm{off}}_j(t_j)=0$ in ${b_j}/{r_j} \leq t_j \leq T$. Second, by letting $b_j/r_j \leq u_j < T$,  we have the expected cost shown as \vspace{-3mm}
\beq
&\!\!\!\!\!\!\! \E[F_j(t_j)] = \int_0^{\frac{b_j}{r_j}} (r_j t_j + b_j)  p'^{\textrm{off}}_j(t_j) d t_j & \nonumber\\
&\!\!\!\!\! +   \int_{\frac{b_j}{r_j}}^{u_j}  (r_j t_j + b_j)  p'^{\textrm{off}}_j(t_j)  d t_j +  \int_{u_j}^T  r_j u_j p'^{\textrm{off}}_j (t_j) d t_j   = \frac{b_j e}{e-1}. \;\;\;\;& \label{eq:expectedcost2int}
\eeq \vspace{-3mm}\\
The second and third terms in \eqref{eq:expectedcost2int} become zero since $p'^{\textrm{off}}_j (t_j) =0$ in ${b_j}/{r_j} \leq t_j \leq T$. By using Definition \ref{eq:cr} and the optimal cost given by \eqref{eq:offlinecost}, the expected competitive ratio of ROA is $\kappa ={e}/{(e-1)}$. As a result, for an arbitrary energy arrival, an ROA provides the OFF time of SBS  that can have the expected cost of ${e}/{(e-1)}$ times of the minimum cost of the problem~\eqref{problem3}. Also, while the ROA has the optimal competitive ratio,  the solutions found by the online algorithms are suboptimal as shown in the definition of competitive ratio \cite{albers2010energy, karlin1994competitive}.  In fact, given uncertainty of energy harvesting, it is challenging to find the optimal solution of problems.

Then, we can derive the average OFF time period of each SBS when the ROA is used to solve problem \eqref{problem2} in the worst case that yields the optimal competitive ratio. 
\begin{theorem}\label{theorem:time}
The expected OFF time period of the SBS is $T-\frac{1}{e-1}\frac{b_j}{r_j}$. 
\end{theorem}\vspace{-3mm}
\begin{proof}
SBS $j$ is turned OFF at time $t_j$, so the OFF time period becomes $T-t_j$.  Therefore, the expected OFF time period within period $T$ is given by $\int_0^{T} (T-t_j)   p'^{\textrm{off}}_j(t_j) d t_j = \int_0^{b_j/r_j} (T-t_j)   \frac{r_j}{b_j} \frac{ e^{({r_j}/{b_j}) t_j}}{e-1} d t_j =  T-\frac{1}{e-1}\frac{b_j}{r_j}.$
\end{proof}
\noindent 
In the classical ski-rental problem, the skiing period is not determined by $T$; thus, the average buying time period cannot  be derived.  However, in our problem, by using a given period $T$, the average OFF time period can be derived.  The result  shows how $b_j$ and  $r_j$ affect the OFF time period.  From the result, if the cost of using the MBS, $b_j$, becomes inexpensive, the OFF time period is longer.  Also, if the cost of using SBS, $r_j$, is decreasing, then the OFF time is reduced, and  the SBSs can be turned ON for a longer time. 

Next, we discuss the case that the ROA solves the original problem \eqref{problem1}. 
Due to the difficulty of theoretical analysis in problem \eqref{problem1}, we numerically evaluate the empirical competitive ratio of the ROA with respect to the problem \eqref{problem1} throughout simulations. 
Furthermore, we carry out simulations to evaluate the OFF time when the ROA is used to solve problem \eqref{problem1} in Section~\ref{sec:numericalresults}. 

\vspace{-3mm}
\section{Simulation Results and Analysis}\label{sec:numericalresults}

\begin{figure}[t]
\centering
\includegraphics[width=7cm]{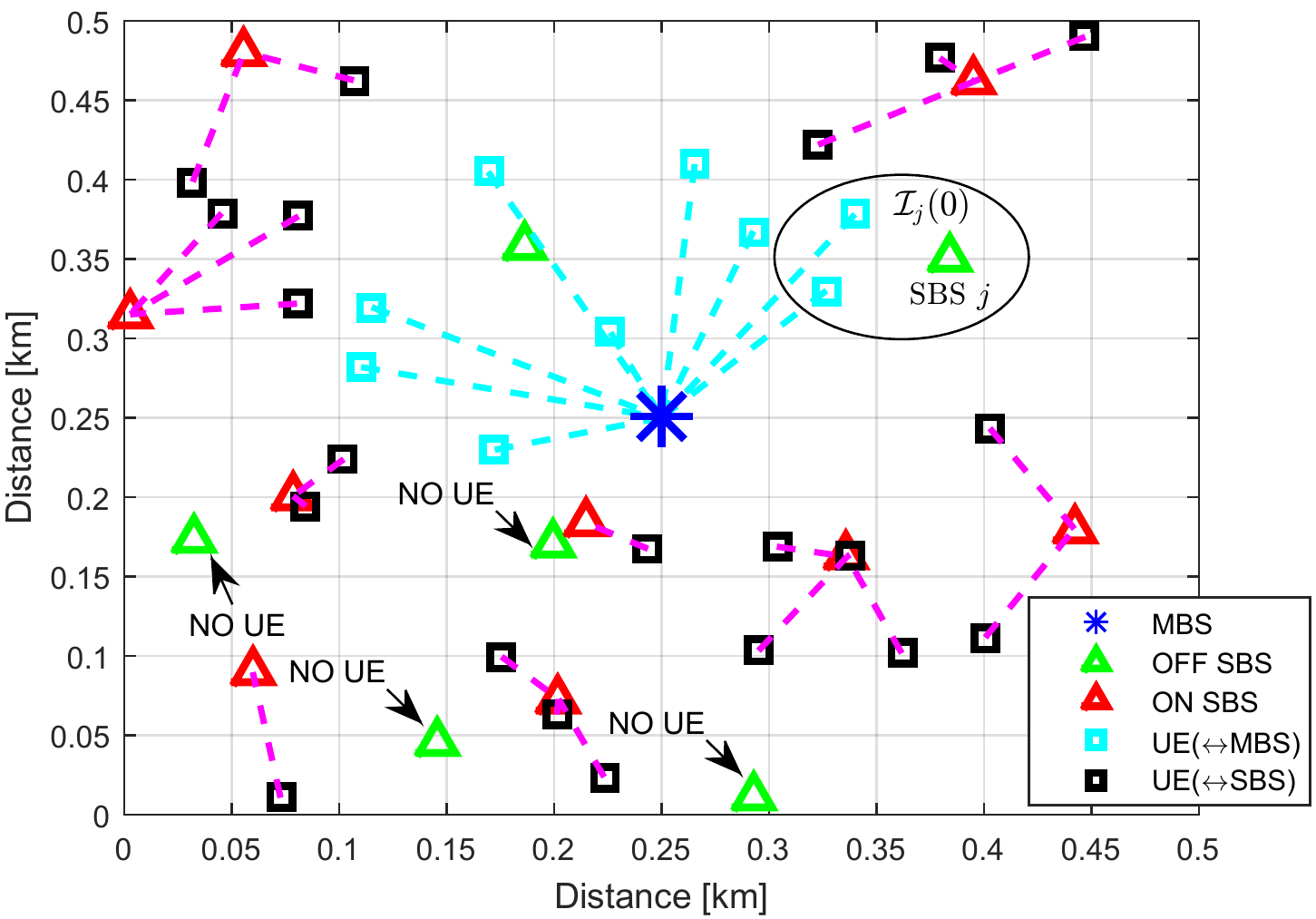}\vspace{-2mm}
\caption{\small Snapshot example of network resulting from the proposed ROA approach.}
\label{fig:snap} 
\end{figure}

For our simulations, the SBSs and UEs are randomly distributed in a $0.5\times0.5$~km$^2$ area with one MBS  located at the center of the area as shown in Fig~\ref{fig:snap}. Statistical results are averaged over a large number of independent simulation runs during time period $2T$ with the parameters in Table~1. Simulations during $2T$  allow a clear observation of the impact of the unused energy in the first period which can be exploited in the next period. In the simulation, all values are updated with the time resolution of  $\epsilon=0.1$~sec. Without loss of generality, during $T\!=\!10~$sec, we assume that energy arrivals per second follow a Poisson process in which energy arrival rate is $20$, and each arrived energy is $0.2~\!$J; for example, it can model a $4~$W solar panel or wind generation having power density of $4~\textrm{W/m$^2$}$\cite{adejumobi2011hybrid}. Also it is assumed that initially  stored energy of SBS $j$ is set to $E_j(0)=60~$J where the maximum capacity of ESS is $E_{\max}\!=\!100~$J. We use $q=0.9$ and $K=10^5~$bits. We compare our online ski rental approach ROA and DOA to the \emph{baseline} approach that turns all SBSs OFF at a certain, pre-determined time $t_j$. 

\begin{table}[t]
\centering
\caption{\small  Simulation parameters}
\label{table:sim}\vspace{-2mm}
\begin{tabular}{|c|c|}
\hline\centering
\footnotesize Notation & Value   \\ \hline
\footnotesize $P^{\textrm{op}}_0$, $P^{\textrm{op}}_j$  & $20~W$, $10~W$       \\ \hline
\footnotesize $P^{\textrm{tx}}_0$,  $P^{\textrm{tx}}_j$ & $33$~dBm, $23$~dBm  \\ \hline
\footnotesize $B_s$, $B_m$            & 10 MHz, 10 MHz      \\ \hline
\footnotesize $M_s$, $M_m$            & 10 users, 50 users  \\ \hline
\footnotesize $\rho^2$                   & {\color{black}-104 dBm   } \\ \hline 
 carrier frequencies  & 2.1 GHz bands  \\ \hline
\end{tabular}\vspace{0mm}
\end{table}

Fig.~\ref{fig:snap} shows a snapshot example for $15$~SBSs, and $30$~UEs at $t=2$ when ROA is used. In Fig.~\ref{fig:snap}, $9$~SBSs are turned ON while $6$ SBSs are turned OFF. Here, user association  is shown as dotted lines between ON SBSs and UEs. From the beginning, four OFF SBSs out of the $6$ OFF SBSs initially stay in the OFF state since they do not have any associated UE as shown in Fig.~\ref{fig:snap}. We can observe that the other two SBSs are turned OFF by the ROA scheduling since the UEs of two OFF SBSs are located near the MBS. In contrast,  most of the ON SBSs are located far from the MBS. In Fig.~\ref{fig:snap}, as UEs in $\cI_j(0)$  are located closer to the MBS, the delay cost of using the MBS, $\phi_0$, decreases. Therefore, the  buy price in \eqref{buy} becomes lower. Thus, as the use of the MBS becomes inexpensive, the SBS tends to buy the MBS resource earlier. Also, as the UEs are located farther from any given SBS $j$, the delay cost of using this SBS, $\phi_j(0)$, will increase.  Thus, the rental  price in \eqref{rental} becomes higher. Since  the use of the SBS becomes more expensive,  the SBS will buy the MBS resource earlier.  

\begin{figure}[t]
\centering 
\includegraphics[width=6cm]{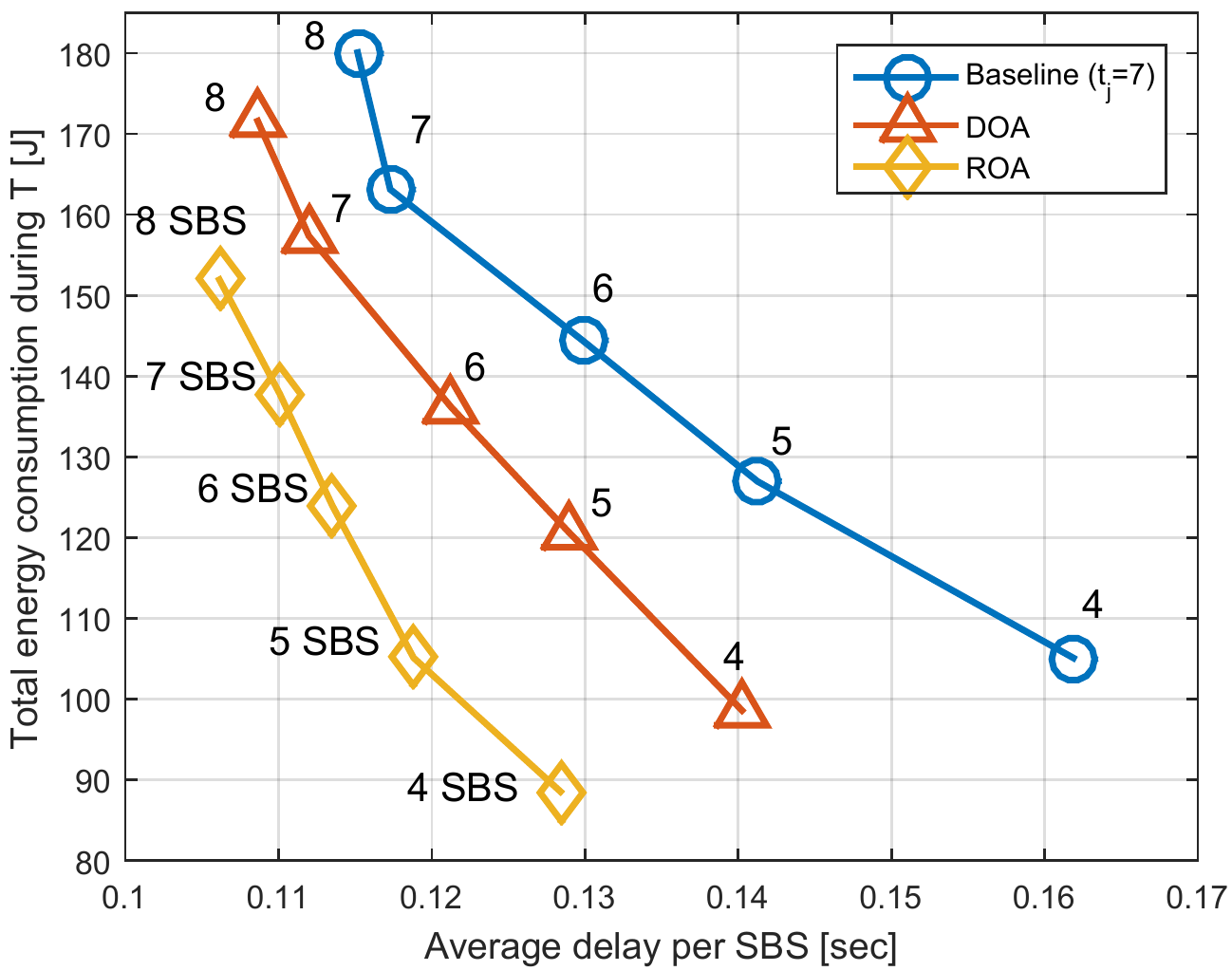}\vspace{-2mm}
\caption{\small {\color{black}Total energy consumption of SBSs} and delay cost per SBS for the ROA, DOA, and baselines.}
\label{fig:powerdelay} 
\end{figure}

Fig.~\ref{fig:powerdelay}  shows, jointly, {\color{black}the total energy consumption of SBSs} and the average network delay per SBS, for various numbers of SBSs with $15$~UEs, $E_j(0)=30$~J, $\alpha_D=0.05$, $\alpha_P=0.0001$, and $\alpha_B=0.05$. From Fig.~\ref{fig:powerdelay}, we can  see that, for all algorithms, as the network size increases, the delay per SBS will decrease, but the total energy consumption will increase. This is due to the fact that having more SBSs turned ON will enable the network to service users more efficiently, however, this comes with an increase in energy consumption. From Fig.~\ref{fig:powerdelay}, we can clearly see that ROA reduces both the delay and the energy consumption as compared to the baseline. It is because ROA results the different turned-OFF time of SBSs while all SBSs are turned OFF at the same designated time in the baseline. Thus, it is possible to mitigate interference and enhance network performance when ROA is used. This performance advantage, reaches up to $20.6\%$ reduction in the delay relative to the baseline $t_j\!=\!7$ for a network with  4 SBSs and $15.6\%$ reduction in energy consumption relative to the baseline for a network with 8 SBSs. Finally, compared to the DOA scheme, Fig.~\ref{fig:powerdelay} shows that ROA will reduce the delay of up to $8.4\%$ (for 4 SBSs) and the energy by up to $11.4\%$ (for 8 SBSs).

\begin{figure}[t]
\centering 
\includegraphics[width=6cm]{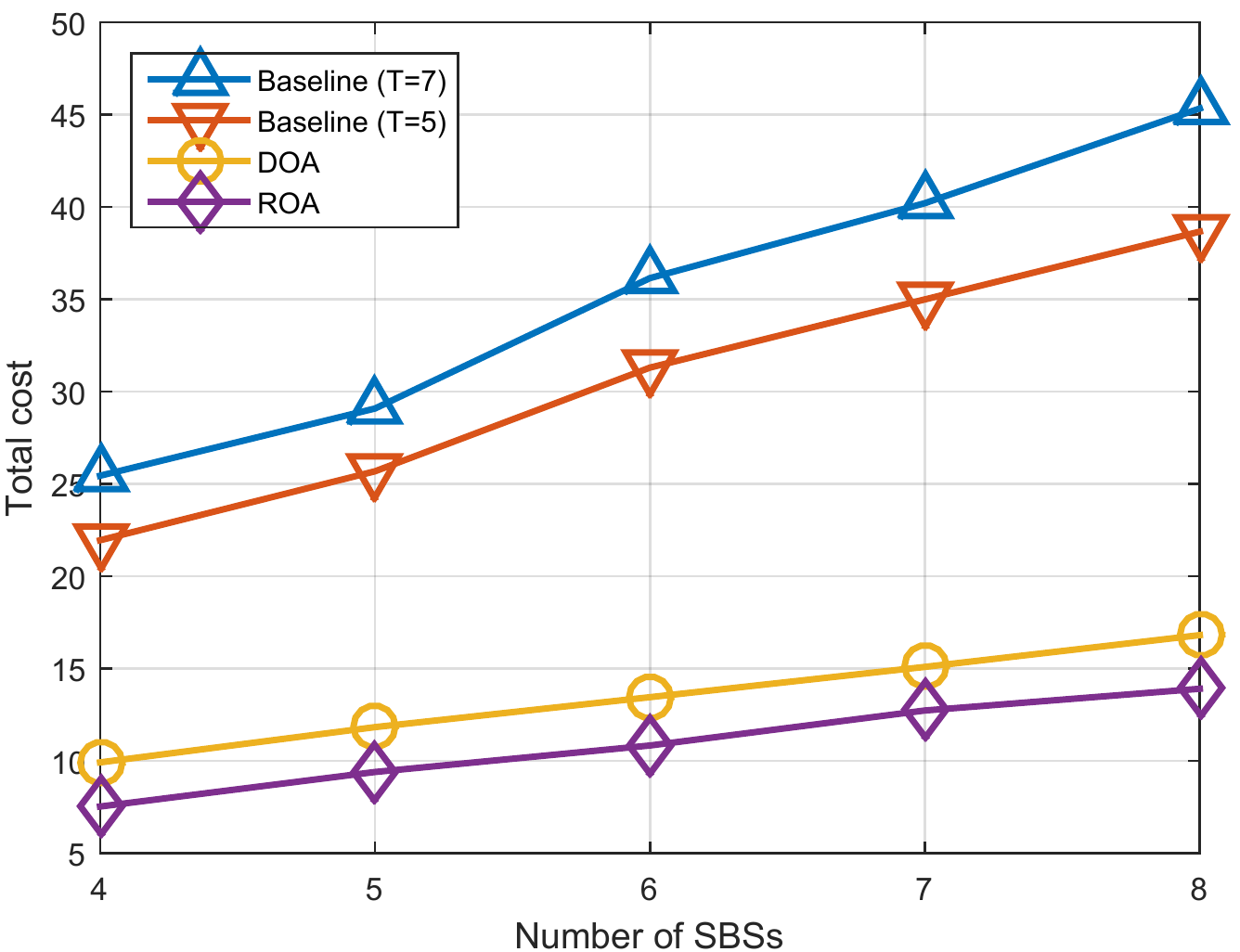}\vspace{-3mm}
\caption{\small \color{black}Comparison of the total network cost when using ROA, DOA, and a baseline.}
\label{fig:obj_sbs} \vspace{-0mm}
\end{figure}

In Fig.~\ref{fig:obj_sbs}, we show the total cost of the network as the network size varies for {\color{black}$30$~UEs},  $\alpha_D=0.05$, $\alpha_P\!=\!0.05$, and {\color{black}$\alpha_B=0.05$}. From Fig.~\ref{fig:obj_sbs}, we can first see that the overall cost of the network given by \eqref{problem1} will increase as the number of SBSs increases. This is mainly due to the fact that increasing the number of SBSs will increase the overall power consumption of the network. Also, the sum of delay of SBSs increases along with the number of SBSs in the network. Fig.~\ref{fig:obj_sbs} shows that the cost increase of the proposed ROA is much slower than the increase of the DOA and the baseline approach. This demonstrates the effectiveness of the proposed approach in maintaining a low network cost. In particular, Fig.~\ref{fig:obj_sbs} shows that, at all network sizes, the proposed online ski rental approach yields  reduction in the overall cost of the network. This performance advantage of ROA reaches up to $69.9\%$ reduction of the average cost for $8$ SBSs compared to the baseline with $t_j=7$. 

\begin{figure}[t]
\centering 
\includegraphics[width=6cm]{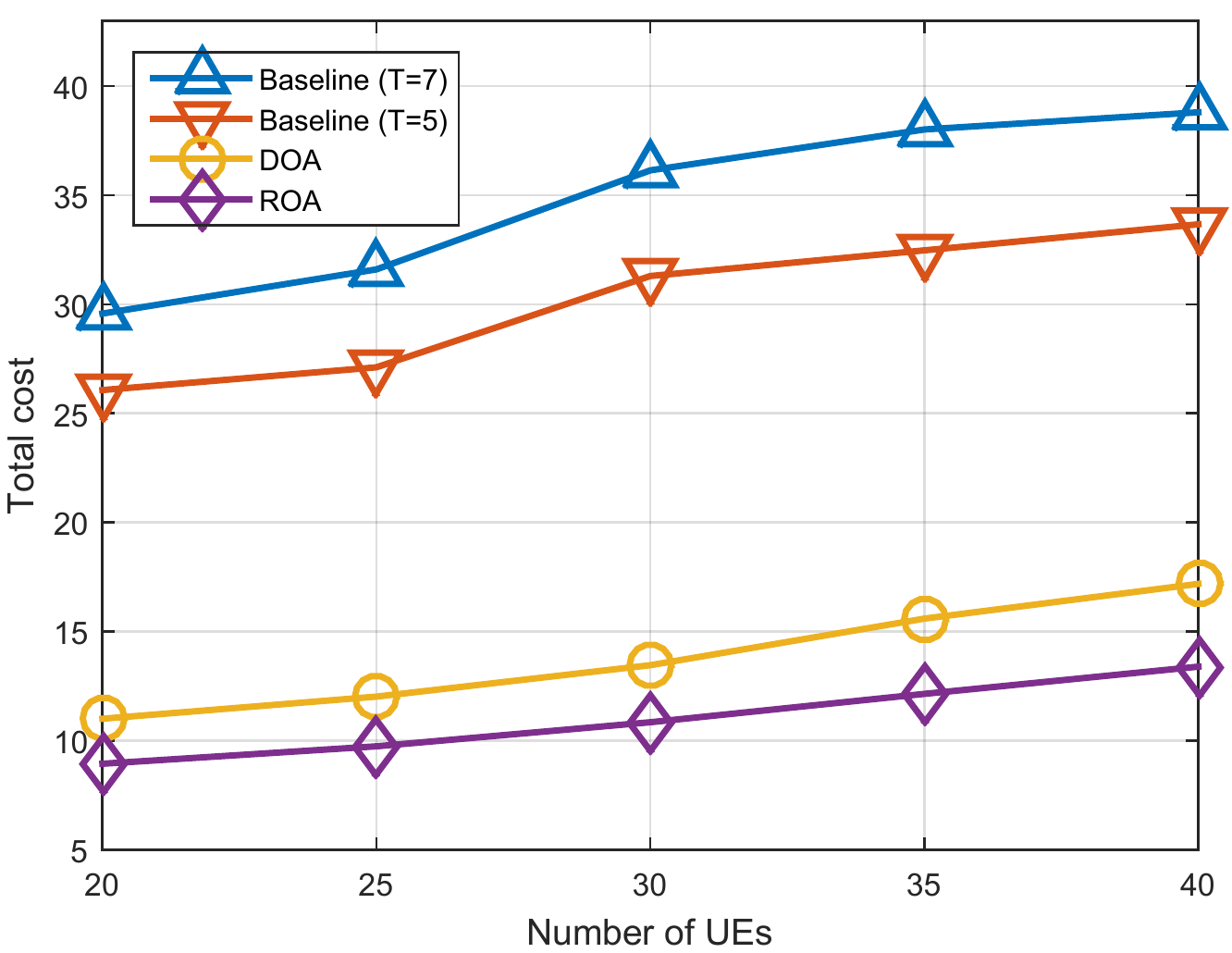}\vspace{-3mm}
\caption{\small \color{black}Comparison of the total network cost with respect to the number of UEs.}
\label{fig:obj_ue} \vspace{-0mm}
\end{figure}

In Fig.~\ref{fig:obj_ue}, the total cost of the network is shown when the number of UEs varies for a network with $6$~SBSs, $\alpha_D=0.05$, $\alpha_P\!=\!0.05$, and {\color{black}$\alpha_B=0.05$}. Fig.~\ref{fig:obj_ue} shows that the total cost of the network increases  along with  the number of UEs. This is because of the fact that increasing the number of UEs will naturally lead to a higher network delay. Nonetheless, we can clearly see that the cost increase of the proposed ROA is  slower than that of the DOA and the baseline approach. This shows that the increase of the overall cost is limited by using the proposed ROA. Fig.~\ref{fig:obj_ue} shows that the performance advantage of ROA can yield a reduction of up to {\color{black}$65.4\%$} of the average cost for {\color{black}$40$~UEs} compared to the baseline $t_j=7$.

\begin{figure}[t]
\centering 
\includegraphics[width=6cm]{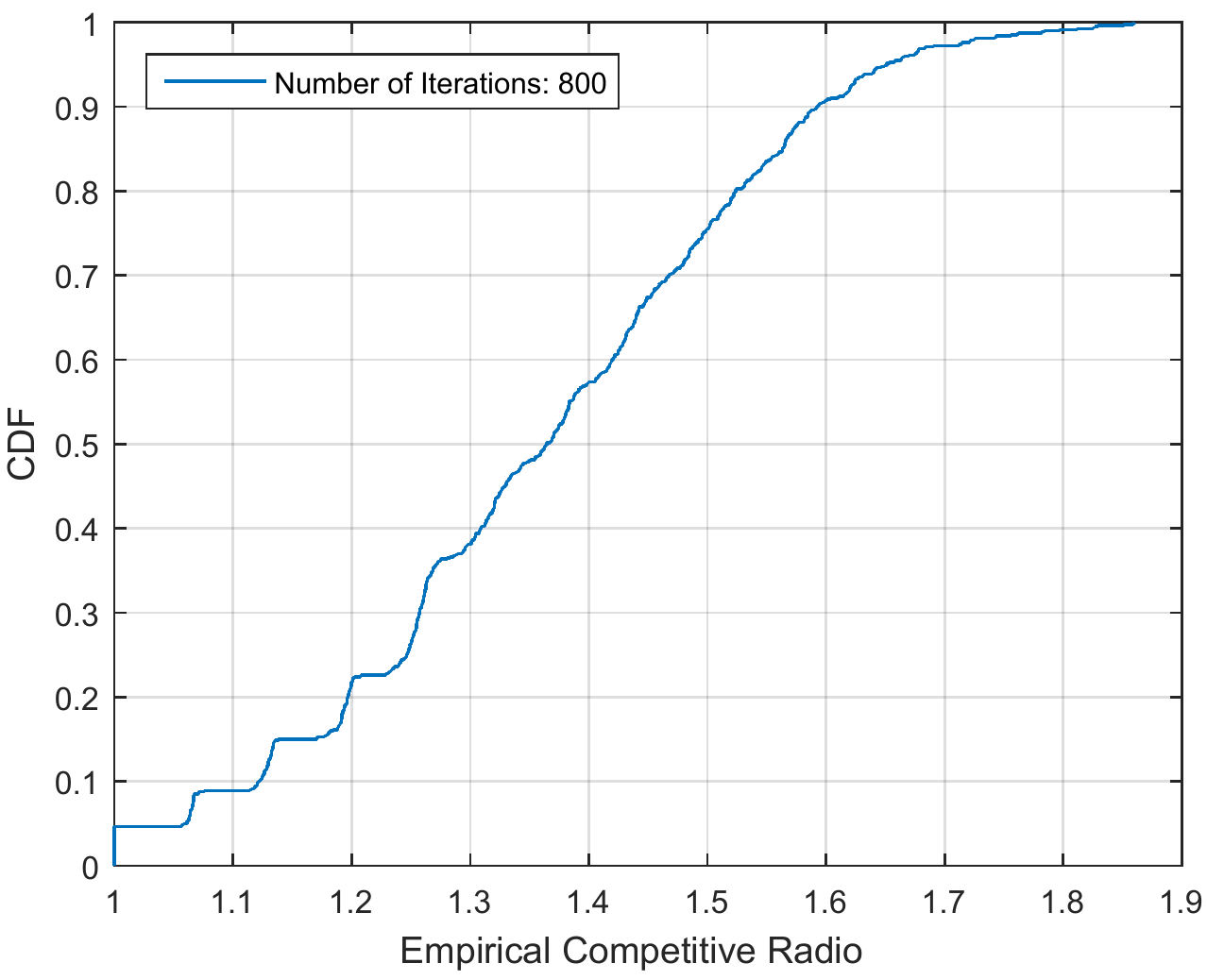}\vspace{-2mm}
\caption{\small Empirical competitive ratio between the total cost of the ROA and the optimal cost.}
\label{fig:cr} \vspace{-0mm}
\end{figure}

Fig.~\ref{fig:cr} shows the empirical competitive ratio  for a network consisting of $3$~SBSs and $15$~UEs with $\alpha_D=0.05$, $\alpha_P\!=\!0.05$, and $\alpha_B=0.05$. To compute empirical competitive ratio, the total cost of the solution resulting from the ROA is divided by the total cost of the offline optimal solution. The optimal cost of each network realization is found by running exhaustive search where all possible OFF times of SBSs are computed.  Since the time complexity of the exhaustive search is $\mathcal{O}\left(\left({T}/{\epsilon}\right)^J\right)$, we reduce the time resolution to $\epsilon=0.2$~sec and run the simulation for one period $T$. We can see that, in 50\% of all iterations, the ROA can yield a total cost that is $1.36$ times that of the offline optimal cost.  Also, over a total of $800$ simulation runs, the empirical competitive ratio in the worst case is shown to be of $1.86$.  Thus, the results show that ROA can effectively choose the OFF time in an online manner. 

\begin{figure}[t]
\centering 
\includegraphics[width=6cm]{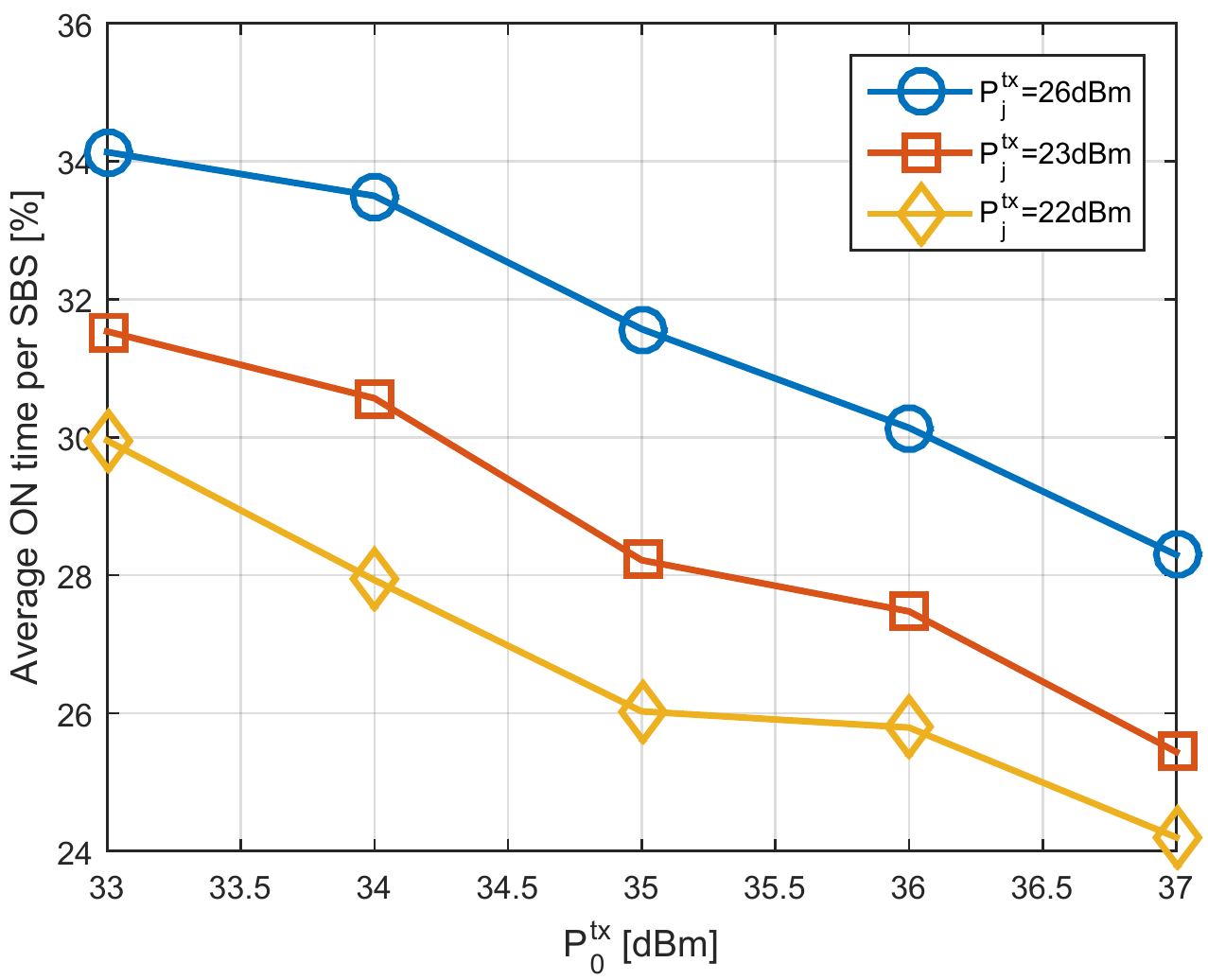}\vspace{-2mm}
\caption{\small Average ON time per SBS with respect to the transmit power of an SBS and the MBS during period $T$.}
\label{fig:ontime_ptx} \vspace{-0mm}
\end{figure}

In Fig.~\ref{fig:ontime_ptx}, the average ON time per SBS  within time period $T$ is shown for  different  transmit powers of an SBS and the MBS with $6$~SBSs, $16$~UEs, $\alpha_D=0.05$, $\alpha_P=0$, and $\alpha_B=0.05$. We compare three different values for the transmit power of an SBS, $P^{\textrm{tx}}_j$: $22$, $23$, and $26$~dBm. If an SBS uses a high $P^{\textrm{tx}}_j$, then the  rent price becomes smaller. As the use of the SBS resource becomes less expensive, the SBS tends to maintain the ON state. This, in turn, results in a longer ON time as shown in Fig.~\ref{fig:ontime_ptx}. For example, the average ON time increase by $16.9\%$ if $P^{\textrm{tx}}_j$ increases from $22$~dBm to $26$~dBm when the MBS uses the transmit power of $37$~dBm. Moreover, if the MBS uses a high $P^{\textrm{tx}}_0$, then the  buy price becomes smaller. As the cost of using the MBS becomes lower, the SBS tends to use the MBS resource. For example, the average ON time per SBS is reduced by $19.2\%$ if $P^{\textrm{tx}}_0$  increases from $33$~dBm to $37$~dBm when the transmit power of $22$~dBm is used by an SBS in a network. 

\begin{figure}[t]
\centering 
\includegraphics[width=6.5cm]{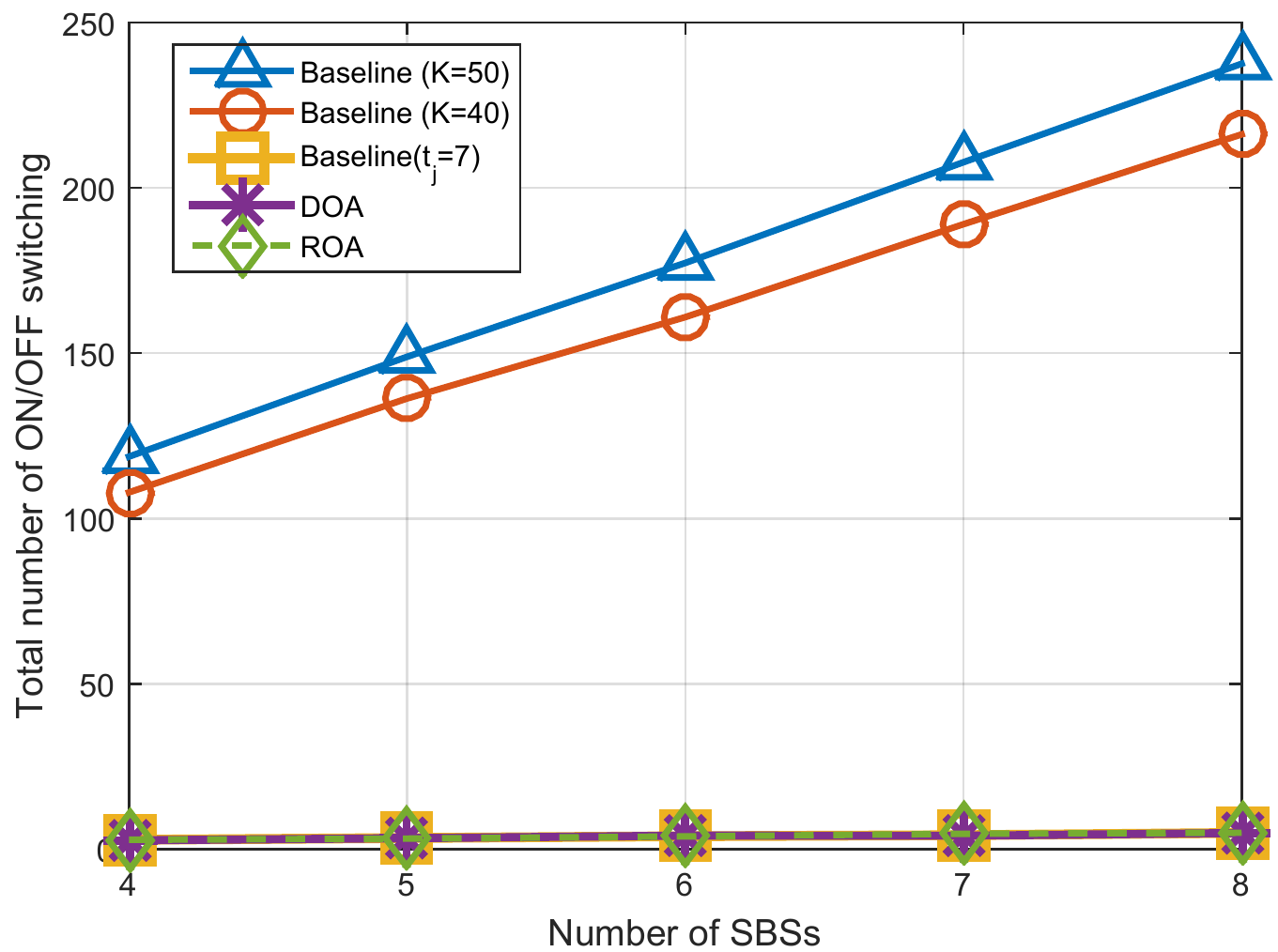}\vspace{-2mm}
\caption{\small The number of ON/OFF switchings of the network during one period $T$.}
\label{fig:switching_sbs} \vspace{-0mm}
\end{figure}

In Fig.~\ref{fig:switching_sbs}, we show the total number of ON/OFF operations within time period $T$ for $16$~UEs, $\alpha_D\!=\!0.05$, $\alpha_P\!=\!0.05$, and $\alpha_B\!=\!0.05$. Here, we consider another \emph{baseline} approach that turns an SBS ON if and only if the percentage of charged energy in storage is greater than a threshold $K$.   For example, we set $K=40$ or $50$ such that an SBS maintains its ESS half-charged. We first present two baselines in which an SBS is turned ON if $K=40$ and $K=50$, respectively. The  ROA and DOA clearly yield a  lower number of SBS ON/OFF switchings whereas the baseline ($K=40$ or $50$) turns SBSs ON and OFF more frequently. This is mainly due to the fact that the algorithm based on the stored energy will turn ON SBSs that have more than a certain predetermined level of energy. However, the ROA and DOA switch SBSs OFF only once in  period $T$. The baseline ($t_j=7$) also  shows the similar number of ON/OFF switchings compared to DOA. Hence, Fig.~\ref{fig:switching_sbs} shows that the  performance advantage of  ROA reaches up to $97.9\%$ of reduction in the number of ON/OFF switchings when compared to the baseline ($K=50$) in the network consisted of 8 SBSs. 

\begin{figure}[t]
\centering 
\includegraphics[width=6cm]{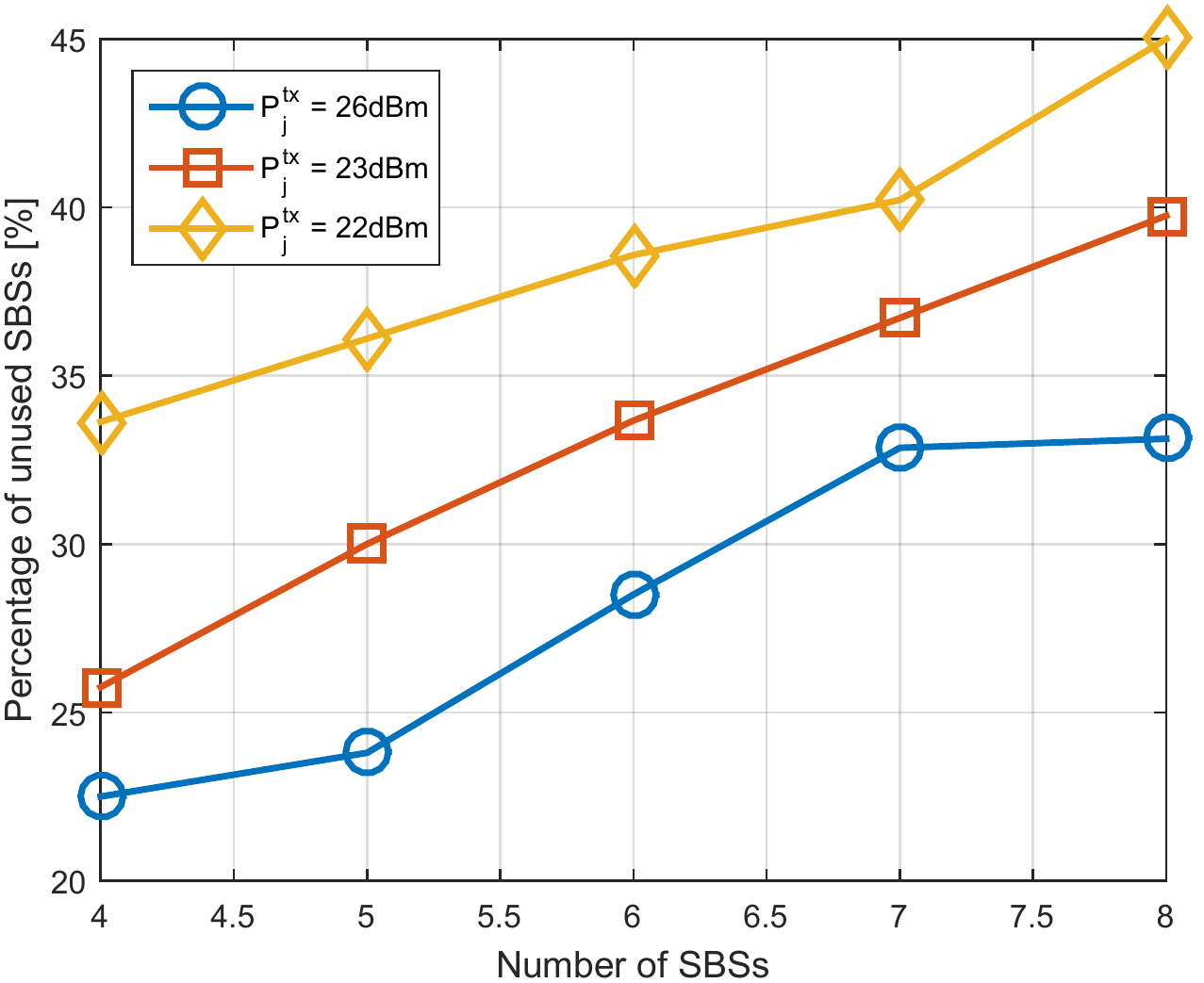}\vspace{-2mm}
\caption{\small Fraction of unused SBSs in the network for the different transmit power.}
\label{fig:unused} \vspace{-0mm}
\end{figure}

In Fig.~\ref{fig:unused}, we show the percentage of unused SBSs for different network sizes. We compare three different values for $P^{\textrm{tx}}_j$: 22, 23, and 26~dBm  for $16$~UEs, $\alpha_D=0.05$, and $\alpha_B=0.05$ while the transmission power of the MBS is fixed to $33$~dBm.  We set $\alpha_P=0$ to observe the changes related to network performance. In Fig.~\ref{fig:unused}, the percentage of unused SBS  decreases as the transmission power of an SBS increases in the network. When the transmission power of an SBS become higher, UEs can receive higher SINR value in \eqref{eq:sinr} than SNR  from the MBS in \eqref{eq:snr}. Thus,  larger number of UEs is connected to SBSs, so it can reduce  the number of unused SBSs. For example, Fig.~\ref{fig:unused} shows that the percentage of unused SBSs is reduced by $33\%$ if the transmission power of an SBS increases from $22$~dBm to $26$~dBm. Also, as the number of SBSs increases, we observe that a higher fraction of SBSs is not used in the network. This is because, as the number of SBSs increases, higher interference will occur thus reducing the SINR at the UEs. In essence, it leads to more UEs that associate with the MBS thus increasing the number of unused SBSs. Indeed, in Fig.~\ref{fig:unused}, we can see that the percentage of unused SBSs increases by $47.2\%$ if the number of SBSs changes from $4$ to $8$ in the network. 

\begin{figure}[t]
\centering 
\includegraphics[width=6cm]{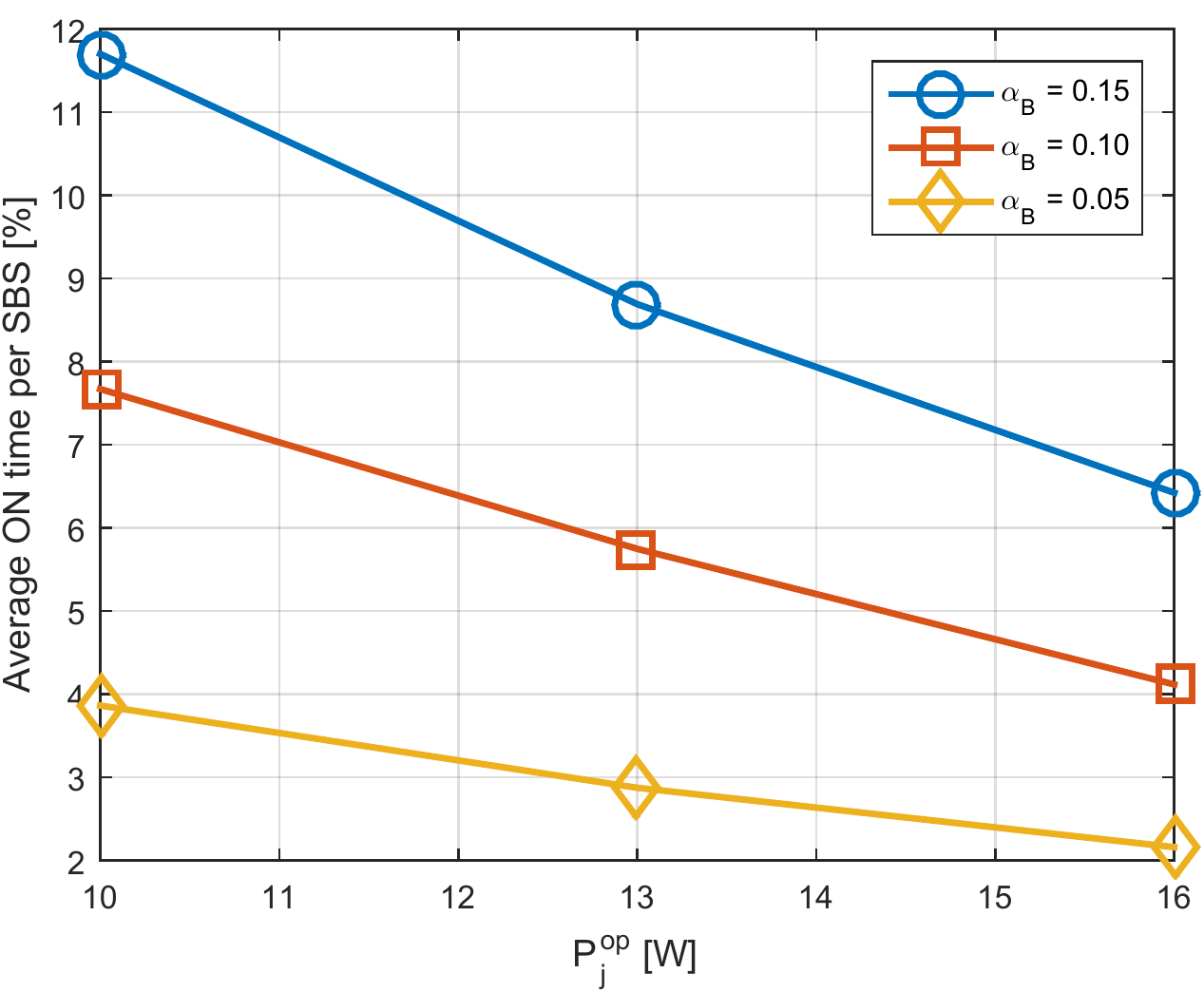}\vspace{-2mm}
\caption{\small Average ON time per SBS for  different operational power of an SBS}
\label{fig:ontime_popsbs_alpha} 
\end{figure}

\begin{figure}[t]
\centering 
\includegraphics[width=6cm]{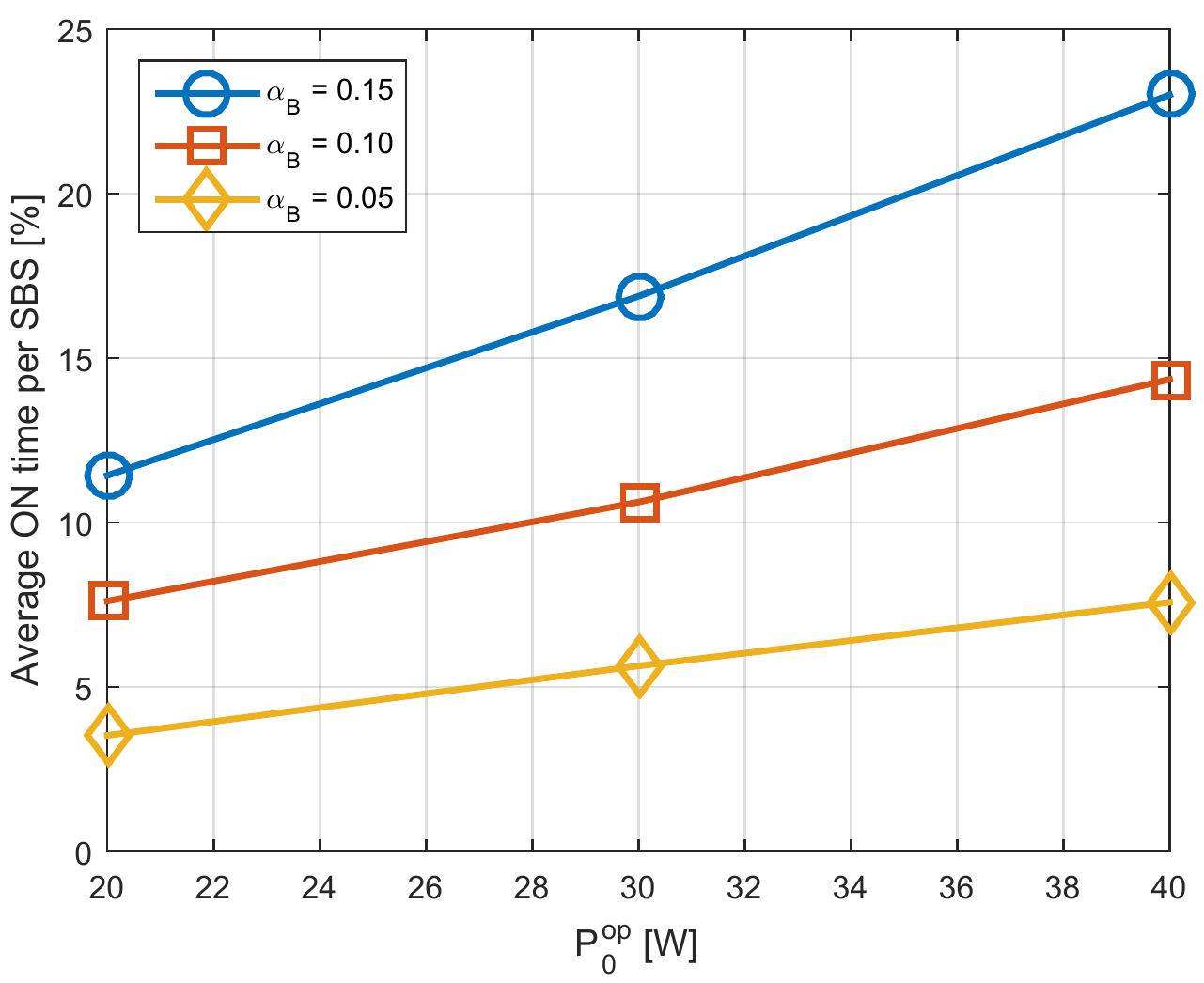}\vspace{-2mm}
\caption{\small Average ON time per SBS for  different operational power of the MBS}
\label{fig:ontime_popmbs_alpha} 
\end{figure}

Figs.~\ref{fig:ontime_popsbs_alpha} and \ref{fig:ontime_popmbs_alpha} show the average ON time per SBS for the different operational power of an SBS $P^{\textrm{op}}_j$ and the MBS $P^{\textrm{op}}_0$, respectively,  with $6$~SBSs, $16$~UEs, and $\alpha_P=0.05$. We set $\alpha_D=0$ to observe the effects from different power consumptions. In Fig.~\ref{fig:ontime_popsbs_alpha}, we observe that the ON time per SBS becomes shorter if an SBS consumes a higher operational power. This is due to the fact that the cost of using an SBS increases with the  power consumption of an SBS. As a result, the rent price becomes higher.  This means that choosing the rent option becomes less affordable, thus resulting in a shorter average ON time per SBS. From Fig.~\ref{fig:ontime_popsbs_alpha}, the average ON time per SBS is shortened by $45~\%$ when the operational power of an SBS is changed from $10$~W to $16$~W when the MBS uses $20$~W. Also, in Fig.~\ref{fig:ontime_popmbs_alpha}, we observe that  the ON time per SBS can be prolonged if the MBS consumes high $P^{\textrm{op}}_0$. This can be explained as follows: if $P^{\textrm{op}}_0$ is high, then the buying price \eqref{buy} becomes higher, so the ON time per SBS becomes longer. The simulation result shows that  the average ON time  increases $2$ times if $P^{\textrm{op}}_0$  increases from $20$~W to $40$~W when an SBS consumes 10~W.  

Furthermore, in Figs.~\ref{fig:ontime_popsbs_alpha} and \ref{fig:ontime_popmbs_alpha}, the average ON time per SBS within time period $T$ is shown for different $\alpha_B$. As $\alpha_B$ becomes larger, a higher buy price will be incurred when an SBS is turned OFF, so the SBS tends to stay in the ON state without buying the MBS resource. This, in turn, results in a longer ON time as shown in Figs.~\ref{fig:ontime_popsbs_alpha}~and~\ref{fig:ontime_popmbs_alpha}. Also, the increase in the ON time is proportional to the  increase of $\alpha_B$. For example, in Fig.~\ref{fig:ontime_popsbs_alpha}, the average ON time increases three folds if $\alpha_B$  increases from $0.05$ to $0.15$ when  $P^{\textrm{op}}_j$ is $10$~W. The same effect can be seen in Fig.~\ref{fig:ontime_popmbs_alpha} where  the average ON time is extended three folds if $\alpha_B$  increases from $0.05$ to $0.15$ when  $P^{\textrm{op}}_0$ is $40$~W. 

\begin{figure}[t]
\centering 
\includegraphics[width=6.5cm]{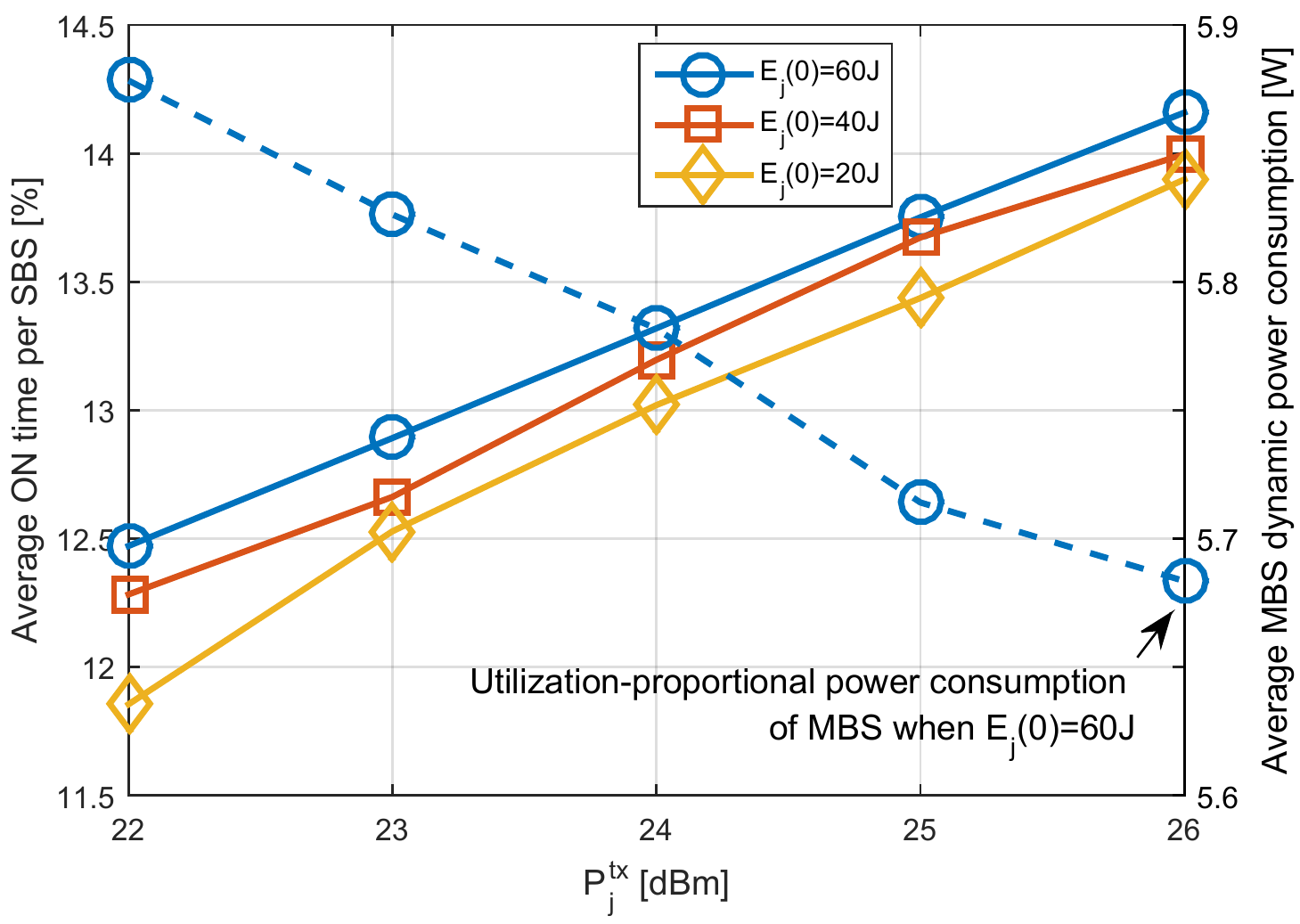}\vspace{-3mm}
\caption{\small Average ON time per SBS for  different harvested energy.}
\label{fig:averageontime_e0} 
\end{figure}

In Fig.~\ref{fig:averageontime_e0}, we show the effect of the initial energy levels on the average ON time for $6$~SBSs, $16$~UEs, $\alpha_D=0.05$, $\alpha_P=0.05$, and  $\alpha_B=0.15$. We compare three different values for $E_j(0)$: 20, 40, and 60~$\!J$ while other parameters related to energy arrival is given equally. As an SBS has high $E_j(0)$, an increase in the average ON time is observed. The result is due to the fact that a high $E_j(0)$ can help an SBS maintain in ON state for a longer period. For instance, the average ON time per SBS increases by $5.1\%$ if $E_j(0)$ increases from $20$~$\!J$ to $60$~$\!J$ when $P^{\textrm{tx}}_j$ is $22$~dBm. Furthermore, we observe that the utilization-proportional power consumption of the MBS is reduced when the ON time per SBS becomes longer. This is because the SBSs will offload  UEs from the MBS. Clearly, the use of self-powered SBSs can  reduce the power consumption of the MBS as shown in the case of $E_j^{\textrm{tx}}(0)=60J$.  

\begin{figure}[t]
\centering 
\includegraphics[width=6cm]{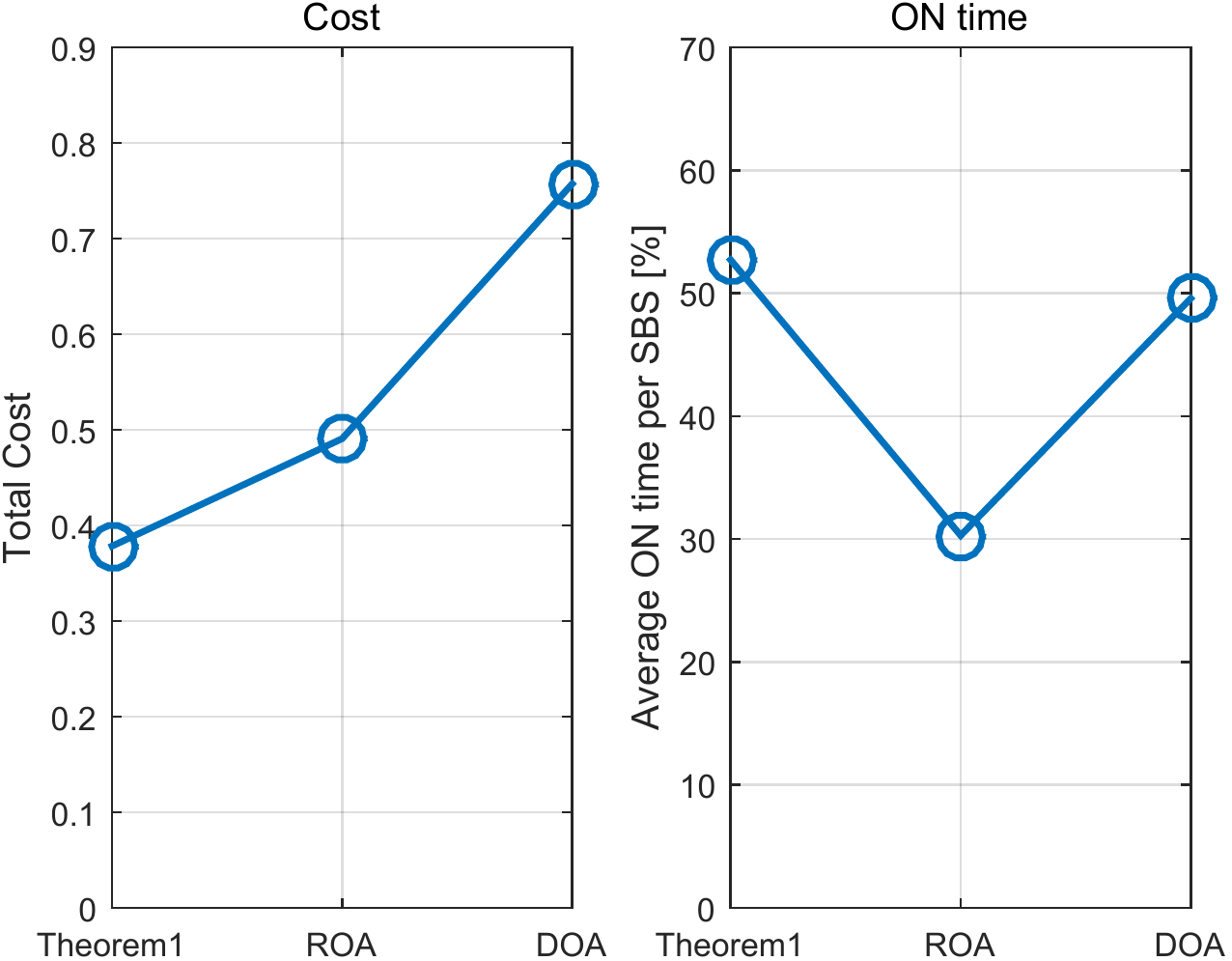}\vspace{-2mm}
\caption{\small Total cost and ON time of an SBS when comparing Theorem~\ref{theorem1},  ROA, and  DOA.}
\label{fig:theorem} 
\end{figure}

In Fig.~\ref{fig:theorem}, we investigate the effect of using more information about the dynamics of the rental cost on minimizing the total cost. We compare the update rule \eqref{eq:bart} in Theorem~\ref{theorem1}, DOA, and ROA under an illustrative network example in which the rental price is monotonically decreasing over time. The considered network here consists of $1$~SBS, $1$~MBS, and $10$~UEs, for $\alpha_D=0.05$, $\alpha_P=0.0001$, and $\alpha_B=0.05$. To satisfy \eqref{eq:decreasing_r}, we set the transmission power $P_j^{\textrm{tx}}$ to $23, 25, 27, \textrm{and } 29$~dBm, at the following time instants $t=0, 1,3, \textrm{and } 5$, respectively. Therefore, when $P_j^{\textrm{tx}}$ increases at $t=1,3, \textrm{and } 5$, the delay cost of the SBS  can be reduced; thus, the rental price  decreases. In this environment, we can observe that the derived update rule in \eqref{eq:bart} can reduce the total cost when it is compared to DOA or ROA. This is due to the fact that by using \eqref{eq:bart}, the SBS can use more information on the updated $P_j^{\textrm{tx}}$ to make a better decision  as opposed to DOA and ROA which rely solely on  only  information. The SBS following \eqref{eq:bart} can dynamically update its decisions based on the decreasing rental cost, so it is possible to have a longer ON time than the DOA as shown in Fig.~\ref{fig:theorem}. For the considered network example, by using \eqref{eq:bart}, SBS will not need to  buy the MBS resourse whereas the DOA uses the SBS resource and also buy the MBS resource. Thus, Theorem~\ref{theorem1} results in the smaller total cost compared to the DOA in the example. Also, ROA can reduce the total cost than the DOA since ROA uses the SBS resource for a short period and chooses to buy the MBS resource earlier. Our example illustrates that the ROA yields a lower cost than the DOA but a higher cost than Theorem~\ref{theorem1}. However, clearly, by using the ROA, the approximation yields a reasonably good solution, which does not require any full information on the dynamic parameters of the system. 

\vspace{-4mm}
\section{Conclusion}\label{sec:conclusion}

In this paper, we have proposed a novel approach to optimize the ON/OFF schedule of self-powered SBSs. We have formulated the  problem that minimizing network operational costs during a period. Also, the problem is approximated as an online ski rental problem which enables the network to operate effectively in the presence of energy harvesting uncertainty. To solve this  online problem, we have proposed deterministic and randomized online algorithm that is shown to achieve the optimal competitive ratio for the approximated problem. Indeed, we have shown that by using the proposed  ROA, each SBS can autonomously  decide on its ON time without knowing any prior information on future energy arrivals. Simulation results have shown that  the proposed ROA can achieve an empirical competitive ratio of $1.86$, thus showing that ROA can effectively choose the OFF time in an online manner. The results have also shown that both delay and the ON/OFF switching overhead are significantly reduced when one adopts the online ski rental approach.

\vspace{-5mm}
\appendix[Proof of Theorem~\ref{theorem1}] 

Given definitions of $r_{(v)}$ and $t_{(v)}$, we determine $\bar{t}$ such that the accumulated cost up to  time $\bar{t}$ equals to the cost of using the MBS $b_j$; thus, $\bar{t}$ satisfies \vspace{-2mm}
\beq\label{eq:equalcost}
&&r_{(1)}t_{(1)}+r_{(2)}(t_{(2)}-t_{(1)})+\cdots\nonumber\\
&&+r_{(v-1)}(t_{(v-1)} -t_{(v-2)} )+ r_{(v)} (\bar{t} -t_{(v-1)} ) = b_j. \qquad
\eeq\vspace{-5mm}\\
At time $t_{(0)}$, the initial SBS's OFF time can be given by $\bar{t}={b_j}/{r_{(1)}}$. At time $t_{(v-1)}$, $v \geq 2$, the SBS's OFF time can be updated by $\bar{t}=\frac{b_j}{r_{(v)}} -  \frac{1}{r_{(v)}} \sum_{v'=1}^{v-1} t_{(v')} (r_{(v')}-r_{(v'+1)}).$ In the algorithm, an SBS determines the OFF time $\bar{t}$  at the beginning, e.g., $t_{(0)}=0$. 
Since the cost is updated from $r_{(v-1)}$ to $r_{(v)}$ at each moment $t_{(v-1)}$, $v \geq 2$, the SBS newly update the OFF time $\bar{t}$ by using \eqref{eq:bart}. 

When the previous OFF time $\bar{t}_{\textrm{old}}$ is determined at $t_{(v-2)}$ with $r_{(v-1)}$,  $\bar{t}$ is updated at $t_{(v-1)}$ with $r_{(v)}$. Then, $\bar{t}$ is shown as \eqref{eq:bart}, and $\bar{t}_{\textrm{old}}$ is given by $\frac{b_j}{r_{(v-1)}} -  \frac{1}{r_{(v-1)}} \sum_{v'=1}^{v-2} t_{(v')} (r_{(v')}-r_{(v'+1)})$. If $\bar{t}_{\textrm{old}} \leq t_{(v-1)}$, the SBS is turned OFF at $\bar{t}_{\textrm{old}}$. Therefore, $\bar{t}_{\textrm{old}} > t_{(v-1)}$ is required so that an SBS is in the ON state at $t_{(v-1)}$. By using two given conditions, $r_{(v-1)}>r_{(v)}$ and $\bar{t}_{\textrm{old}} > t_{(v-1)}$, the inequality  $\bar{t} = \frac{1}{r_{(v)}} \left(r_{(v-1)} \bar{t}_{\textrm{old}} - t_{(v-1)} (r_{(v-1)}-r_{(v)})    \right) >\bar{t}_{\textrm{old}}$ holds. Hence, the updated OFF time $\bar{t}$ is later than the previous OFF time $\bar{t}_{\textrm{old}}$ if $r_{(v-1)} > r_{(v)}$. 

For an arbitrary $v \geq 2$, the OFF time of SBS $j$ can be determined at time $t_{(v-1)}$ by \eqref{eq:bart}. 
Also, the energy of the SBS can be depleted at time $u$ where $u \geq t_{(v-1)}$. To derive the competitive ratio, we show the total cost of the algorithm and the optimal cost, respectively. If $t_{(v-1)} \leq u < \bar{t}$, then the total cost of the problem in \eqref{problem1} is given by\vspace{-2mm}
\beq\label{eq:alg1}
\beta_{\textrm{ALG}}(u) = \sum_{v'=1}^{v-1} t_{(v')} (r_{(v')}-r_{(v'+1)}) + r_{(v)} u.
\eeq\vspace{-4mm}\\
The optimal cost $\beta_{\textrm{OPT}} (u)$ can be calculated by assuming an offline scenario where energy arrival information over the entire  period is given.  Thus, the amount of stored energy at each moment   becomes  known information. In this case, we can find that $\beta_{\textrm{OPT}}(u)$ is the same as \eqref{eq:alg1}. 
Also, if $\bar{t} \leq u$, then the total cost is given by\vspace{-2mm}
\beq\label{eq:alg2}\nonumber
\beta_{\textrm{ALG}}(u) = \sum_{v'=1}^{v-1} t_{(v')} (r_{(v')}-r_{(v'+1)}) + r_{(v)} \bar{t} + b_j.
\eeq\vspace{-4mm}\\
However, the offline optimal cost is given by $\beta_{\textrm{OPT}}(u)=b_j$. Therefore, the worst-case competitive ratio given by \eqref{eq:defcr} becomes $2$ in the case of $\bar{t} \leq u$ since $\beta_{\textrm{ALG}}(u)$ can be two times greater than $\beta_{\textrm{OPT}}(u)$ due to \eqref{eq:equalcost}. 

\vspace{-4mm}
\bibliographystyle{IEEEtran}

\end{document}

%% file: mymath.tex
\def\cB{\mathcal{B}}

\def\cI{\mathcal{I}}
\def\cJ{\mathcal{J}}

\usepackage[only,tensf,]{rawfonts}

\newcommand{\tx}[1]{\text{tx}(l)}
\newcommand{\rx}[1]{\text{rx}(l)}

\newcommand{\separator}{
  \begin{center}
    \rule{\columnwidth}{0.3mm}
  \end{center}
}

\newcommand{\bi}{\begin{itemize}}
\newcommand{\ei}{\end{itemize}}

 \newtheorem{theorem}{Theorem}
 \newtheorem{proposition}{Proposition}
 \newtheorem{definition}{Definition}
 \newtheorem{remark}{Remark}

\newcommand{\be}{\begin{equation}}
\newcommand{\ee}{\end{equation}}
\newcommand{\bea}{\begin{equation*}}
\newcommand{\eea}{\end{equation*}}
\newcommand{\beq}{\begin{eqnarray}}
\newcommand{\eeq}{\end{eqnarray}}
\newcommand{\beqa}{\begin{eqnarray*}}
\newcommand{\eeqa}{\end{eqnarray*}}

\def\E{\mathbb{E}}

\def\log{\mbox{log}}

\renewcommand\floatsep{0mm}
\renewcommand\textfloatsep{2mm}